\documentclass[12pt]{article}
\usepackage{a4wide,epsfig,amsmath,amssymb,cite,scalefnt,graphicx}
\numberwithin{equation}{section}
\usepackage{axodraw4j}
\usepackage{pstricks}
\usepackage{color}
\usepackage{changepage}
\usepackage{tabularx}
\usepackage{array}
\usepackage{epstopdf}

\def\Ttil{\tilde T}
\def\npb{{\it{Nucl.\ Phys.}\ }{\bf B}}
\def\plb{{{\it Phys.\ Lett.}\ }{ \bf B}}
\def\jhep{{\it JHEP}\ }
\def\be{\begin{equation}}
\def\ee{\end{equation}}
\def\nn{\nonumber\\}

\def\tr{\hbox{tr}}

\def\prd{{{\it Phys.\ Rev.}\ }{\bf D}}

\def\Ybar{\overline Y}

\def\thetabar{{\overline\theta}}

\def\fbar{\overline f}
\def\Fbar{\overline F}
\def\Lambdabar{\overline \Lambda}

\def\Tr{\mathop{\rm Tr}}

\def\Ttil{\tilde T}

\def\frakk[#1#2{{{#1}\over{#2}}}

\def\lambdabar{\overline\lambda}

\def\Dbar{{\overline D}}

\def\Fbar{\overline F}

\def\Ttil{\tilde T}

\def\Ncal{{\cal N}}

\def\pa{\partial}
\def\pabar{\overline\partial}

\input amssym.def
\input amssym
\def\npb{Nucl. Phys. B}
\def\plb{Phys. Lett. B}
\def\pa{\partial}
\def\be{\begin{equation}}
\def\ee{\end{equation}}
\def\nn{\nonumber\\}

\def\tr{\hbox{tr}}

\def \tr{{\rm tr }}
\def \ta{{\tilde a}}

\def\ta{\tilde a}

\def\cirk{\,{\raise1pt \hbox{${\scriptscriptstyle \circ}$}}\,}

\def \olr{{\raise6.5pt\hbox{$\leftrightarrow  \! \! \! \! \!$}}}
\newcommand{\kslash}{k \! \! \! /}

\font\ninerm=cmr9 \font\ninesy=cmsy9
\font\eightrm=cmr8 \font\sixrm=cmr6
\font\eighti=cmmi8 \font\sixi=cmmi6
\font\eightsy=cmsy8 \font\sixsy=cmsy6
\font\eightbf=cmbx8 \font\sixbf=cmbx6
\font\eightit=cmti8
\def\eightpoint{\def\rm{\fam0\eightrm}
  \textfont0=\eightrm \scriptfont0=\sixrm \scriptscriptfont0=\fiverm
  \textfont1=\eighti  \scriptfont1=\sixi  \scriptscriptfont1=\fivei
  \textfont2=\eightsy \scriptfont2=\sixsy \scriptscriptfont2=\fivesy
  \textfont3=\tenex   \scriptfont3=\tenex \scriptscriptfont3=\tenex
  \textfont\itfam=\eightit  \def\it{\fam\itfam\eightit}%
  \textfont\bffam=\eightbf  \scriptfont\bffam=\sixbf
  \scriptscriptfont\bffam=\fivebf  \def\bf{\fam\bffam\eightbf}%
  \normalbaselineskip=9pt
  \setbox\strutbox=\hbox{\vrule height7pt depth2pt width0pt}%
  \let\big=\eightbig  \normalbaselines\rm}
\catcode`@=11 %
\def\eightbig#1{{\hbox{$\textfont0=\ninerm\textfont2=\ninesy
  \left#1\vbox to6.5pt{}\right.\n@@space$}}}
\def\vfootnote#1{\insert\footins\bgroup\eightpoint
  \interlinepenalty=\interfootnotelinepenalty
  \splittopskip=\ht\strutbox %
  \splitmaxdepth=\dp\strutbox %
  \leftskip=0pt \rightskip=0pt \spaceskip=0pt \xspaceskip=0pt
  \textindent{#1}\footstrut\futurelet\next\fo@t}
\catcode`@=12 %
\def\today{\number\day\ \ifcase\month\or January\or February\or March\or
April\or May\or June\or July\or
August\or September\or October\or November\or December\fi, \number\year}

\input epsf

\begin{document}

\begin{titlepage}
\begin{flushright}
LTH1097\\
\end{flushright}
\date{}
\vspace*{3mm}

\begin{center}
{\Huge The $a$-function for $\Ncal=2$ supersymmetric gauge theories in three dimensions}\\[12mm]
{\bf J.A.~Gracey\footnote{{\tt gracey@liverpool.ac.uk}}, I.~Jack\footnote{{\tt dij@liverpool.ac.uk}} and
C.~Poole\footnote{{\tt c.poole@liverpool.ac.uk}}}\\

\vspace{5mm}
Dept. of Mathematical Sciences,
University of Liverpool, Liverpool L69 3BX, UK\\
\vspace{10mm}

{\bf Y.~Schr\"oder\footnote{{\tt yschroeder@ubiobio.cl }}}\\

\vspace{5mm}
Grupo de Fisica de Altas Energias, Universidad del Bio-Bio,
Casilla 447, Chillan, Chile

\end{center}

\vspace{3mm}

\begin{abstract}
Recently, the existence of a candidate $a$-function for renormalisable theories in three dimensions was demonstrated for a general theory at leading order and for a scalar-fermion theory  at next-to-leading order.
Here we extend this work by constructing the $a$-function at next-to-leading order for an $\Ncal=2$  supersymmetric Chern-Simons theory. This increase in precision for the $a$-function necessitated the evaluation of the underlying renormalization-group functions at {\em four} loops.
\end{abstract}

\vfill

\end{titlepage}

\section{Introduction}

Following Cardy's suggestion\cite{Cardy} that Zamolodchikov's two-dimensional $c$-theorem\cite{Zam} might have an analogue in four dimensions, considerable progress has been made in proving the so-called $a$-theorem in even dimensions \cite{KS,KS1,Luty,ElvangST,ElvangST1,Analog,Analog1,OsbJacnew,Weyl,GrinsteinCKA, GrinsteinCKA1,GrinsteinCKA2,GrinsteinCKA3} (for a review see Ref.~\cite{nak}). The Weyl anomaly played a central role in the derivation of the $c$-theorem and the $a$-theorem in even dimensions, and therefore it seems unlikely that the $a$-theorem could be extended to odd dimensions where there is no Weyl anomaly. An alternative candidate for a function which evolves monotonically along renormalisation group (RG) flows in odd dimensions is the so-called
 $F$-function\cite{jaff,klebb,kleba,klebc}. This is the Euclidean path integral of the theory (or ``free energy'') conformally mapped (in the case of three dimensions) to $S^3$. It has been shown to increase between UV and IR fixed points for a variety of theories. However, an additional important feature of the $a$-function in even dimensions is the gradient flow property; for theories with couplings $g^I$ and corresponding RG $\beta$-functions 
$\beta^I$, it satisfies 
\be \pa_I a \equiv \frac{\pa}{\pa g^I}a=T_{IJ}\beta^J\,
\label{grad}
\ee
for a function  $T_{IJ}$. A crucial consequence of Eq.~\eqref{grad} is that we then have
\be
\mu \frac{d}{d\mu} a=\beta^I\frac{\pa}{\pa g^I} a=G_{IJ} \beta^I\beta^J
\ee
where $G_{IJ}=T_{(IJ)}$, $(IJ)$ denoting symmetrisation. The $a$-theorem then follows immediately if  $G_{IJ}$ is positive definite. There is, however,  so far no evidence that $F$ possesses the gradient flow property except in simple cases at leading order, where its existence is trivial in the sense that no conditions are imposed on the $\beta$-function coefficients. 

Accordingly a different approach has recently been taken\cite{JJP,JPa} in which a function with the 
gradient flow property Eq.~\eqref{grad} has been constructed order by order, using as a starting point the $\beta$-functions for a range of three-dimensional theories.  The method was essentially that employed in four dimensions in the classic work of Ref.~\cite{Wallace}.
Initially\cite{JJP} the leading-order (two-loop) $\beta$-functions computed by Avdeev et al in Refs.~\cite{kaza,kazb} were used to construct a solution of Eq.~\eqref{grad} for abelian and non-abelian (for the case $SU(n)$) Chern-Simons theories at leading order. Moreover the ``metric'' $G_{IJ}$ was indeed positive definite at this order.
The Yukawa and scalar couplings in these theories were of a restricted form. However, it was then shown\cite{JJP,JPa} that the results
extended at leading order to completely general abelian Chern-Simons theories (and there was no reason to doubt that the extension to the non-abelian case would be fairly immediate). The extension to next-to-leading order involves the four-loop $\beta$-functions (recall that there are no divergences at odd loop orders in odd dimensions). The four-loop Yukawa $\beta$-function for a general (ungauged) fermion/scalar theory was therefore computed\cite{JPa} and it was shown that the definition of the $a$-function in Eq.~\eqref{grad} could be extended to this order as well. In the general gauged case at leading order, and in the ungauged case at next-to-leading order, Eq.~\eqref{grad} imposes stringent conditions on the $\beta$-function coefficients. Clearly (in the absence of a general proof of the gradient flow property) the next natural step would be to extend the calculation for a general gauge theory to next-to-leading order. The $\beta$-function computation for a general theory at four loops would be very involved; however, the supersymmetric case is much more tractable and consequently we consider here the case of $\Ncal=2$ supersymmetry. Here we can avail ourselves of the superspace formalism to simplify the calculations and furthermore as a consequence of the nonrenormalisation theorem\cite{GRS}, the Yukawa $\beta$ function is determined by the chiral field anomalous dimension. Another motivation for consideration of the supersymmetric case is that the $F$-theorem has mostly been studied in this context and therefore a comparison might be facilitated. In order to check the validity of Eq.~\eqref{grad}, it is sufficient to compute only the contributions to the anomalous dimension which contain Yukawa couplings. This is because Eq.~\eqref{grad} places constraints on the Yukawa-dependent contributions (which we shall show are satisfied) but not on the Yukawa-independent terms. This is a fortunate situation since it saves us a great deal of arduous computation.

We find that indeed we can construct the $a$-function satisfying Eq.~\eqref{grad} at next to leading order in the case
of a general $\Ncal=2$ supersymmetric Chern-Simons theory. On the one hand there are far fewer constraints (only two, in fact) on the RG coefficients than we found even in the ungauged non-supersymmetric case at next-to-leading order; so that the imposition of supersymmetry itself must guarantee that most of the original constraints are satisfied. On the other hand, one of the remaining constraints is highly non-trivial since it involves a constraint on a Feynman diagram which had not hitherto been computed and which we had to deploy advanced techniques to evaluate. 

The structure of the paper is as follows: in Section 2 we describe the  $\Ncal=2$ Chern-Simons theory and its quantisation, recall the lowest-order (two-loop) result for the anomalous dimension and use it to construct the leading-order term in the $a$-function.  In Section 3 we consider the $a$-function at next-to leading order, and show how its existence imposes consistency conditions on some of the coefficients in the next-to-leading order (four-loop) anomalous dimension. In Section 4 we present our calculation of the Yukawa-dependent terms in the four-loop anomalous dimension (as explained earlier, these are all we need), in particular checking that the consistency conditions are satisfied. We explain in some detail the computation of the particular diagram mentioned above. Finally, some closing remarks are offered in the Conclusion. Details of the superspace conventions and some explicit results for momentum integrals are deferred to appendices.

\section{General procedure and lowest-order results}
 In this section we describe the $\Ncal=2$ Chern-Simons theory and the general framework for our calculations.
We also review the two-loop anomalous dimension calculation and construct the corresponding $a$-function. The action for the theory can be written
 \be
 S=S_{SUSY}+S_{GF}
 \label{lag}
 \ee
 where $S_{SUSY}$ is the usual supersymmetric action\cite{ivanov}
 \begin{align}
 S_{SUSY}=&\int d^3x\int d^4\theta\left(
 k\int_0^1dt\Tr[\Dbar^{\alpha}(
 e^{-tV}D_{\alpha}e^{tV})]+\Phi^j (e^{V_AR_A})^i{}_j\Phi_i\right)\nn
 &+\left(\int d^3x\int d^2\theta \,W(\Phi)
 +\hbox{h.c.}\right).
 \label{ssusy}
 \end{align}
Here $V$ is the vector
 superfield, $\Phi$ the chiral matter superfield and the
 superpotential $W(\Phi)$ (quartic for renormalisability in three dimensions); see Appendix A for our $\Ncal=2$ superspace 
conventions.  We take $W(\Phi)$ to be given by
 \be
 W(\Phi)=\tfrac{1}{4!}Y^{ijkl}\Phi_i\Phi_j\Phi_k\Phi_l.
 \ee
 (We use the
 convention that $\Phi^i=(\Phi_i)^*$, and also denote $\Ybar_{ijkl}=(Y^{ijkl})^*$.)
 We assume a simple gauge group, though we comment later on the extension to
 non-simple groups.
 Gauge invariance requires the gauge coupling $k$ to be
 quantised, so that $2\pi k$ is an integer. The vector superfield $V$ is in the
 adjoint representation, $V=V_AT_A$ where $T_A$ are the generators of the
 fundamental representation, satisfying
 \begin{align}
 [T_A,T_B]=&if_{ABC}T_C,\nn
 \Tr(T_AT_B)=&\delta_{AB}.
 \end{align}
 The chiral superfield can be in a general representation, with gauge
 matrices denoted $R_A$ satisfying
 \begin{align}
 [R_A,R_B]=&if_{ABC}R_C,\nn
 \Tr(R_AR_B)=&T_R\delta_{AB}.
 \label{Tdef}
 \end{align}
 In three dimensions the Yukawa couplings $Y^{ijkl}$ are dimensionless
 and (as mentioned earlier) the theory is renormalisable.
 In Eq.~(\ref{lag}) the gauge-fixing term $S_{GF}$ is given by\cite{penatib}
 \be
 S_{GF}=-\tfrac{k}{2\xi}\int d^3xd^2\theta\,\tr[f\fbar]
 -\tfrac{k}{2\xi}\int d^3xd^2\thetabar\,\tr[f\fbar],
 \label{sgf}
 \ee
 where $\xi$ is a gauge-fixing parameter, and we introduce into the functional integral a corresponding ghost term
 \be
 \int{\cal D}f{\cal D}\fbar\Delta(V)\Delta^{-1}V
 \ee
 with
 \be
 \Delta(V)=\int d\Lambda d \Lambdabar\,\delta(F(V,\Lambda,\Lambdabar)-f)
 \delta(\Fbar(V,\Lambda,\Lambdabar)-\fbar),
 \label{ghostdet}
 \ee
 and $\Fbar=D^2V$, $F=\Dbar^2V$.
 With the gauge-fixing parameter $\xi=0$ this results in a gauge propagator
 \be
 \langle V^A(1)V^B(2)\rangle=-\tfrac1k\tfrac{1}{\pa^2}
 \Dbar^{\alpha}D_{\alpha}\delta^4(\theta_1-\theta_2)\delta^{AB}.
 \ee

 The gauge vertices are obtained by expanding $S_{SUSY}+S_{GF}$ as given by
 Eqs.~(\ref{ssusy}), (\ref{sgf}):
 \begin{align}
 S_{SUSY}+S_{GF}\rightarrow&
 -\tfrac{i}{6}f^{ABC}\int d^3xd^4\theta\Dbar^{\alpha}V^A
 D_{\alpha}V^BV^C\nn
 -&\tfrac{1}{24}f^{ABE}f^{CDE}\int d^3xd^4\theta\Dbar^{\alpha}V^AV^B
 D_{\alpha}V^CV^D+\ldots.
 \end{align}

 The ghost action resulting from Eq.~(\ref{ghostdet}) has the same form
 as in the four-dimensional $\Ncal=1$ case\cite{GGRS,GRS}; we refrain from
quoting it explicitly since we do not need to consider diagrams with ghost propagators.
 Finally the chiral propagator and chiral-gauge vertices are readily obtained
 by expanding Eq.(\ref{ssusy}); the chiral propagator is given by
 \be
 \langle \Phi^i(1)\Phi_j(2)\rangle=-\tfrac{1}{\pa^2}\delta^4(\theta_1-\theta_2)
 \delta^i{}_j.
 \ee
 The regularisation of the theory is effected by replacing
 $V$, $\Phi$, $Y$ by corresponding bare quantities $V_B$, $\Phi_B$, $Y_B$,
 with the bare and renormalised fields related by
\be
 V_B=Z_V^{\tfrac12}V,\quad \Phi_B=Z_{\Phi}^{\tfrac12}\Phi.
 \ee
 Since the Chern-Simons level $k$ is expected to be unrenormalised for a
 generic Chern-Simons theory due to the topological nature of the theory
 (so that $k_B=k$), the only
 $\beta$-functions are those for the superpotential coupling and its conjugate. These will be
 given according to the non-renormalisation theorem\cite{GRS} by
 \be
 \beta_Y^{ijkl}=(\gamma_{\Phi})_m{}^{(i}Y^{jkl)m}, \quad \beta_{\Ybar ijkl}=\Ybar_{m(ijk}(\gamma_{\Phi})_{l)}{}^{m}.
 \label{nonren}
 \ee
 where the anomalous dimension $\gamma_{\Phi}$ is defined by
 \be
 \gamma_{\Phi}=\frac12\mu\frac{d}{d\mu}\ln Z_{\Phi}.
 \ee
Using dimensional regularisation with $d=3-\epsilon$ dimensions, we have
 \be
 Z_{\Phi}=\sum_{L\,\rm{even},m=1\ldots \tfrac{L}{2}}
 \frac{Z_{\Phi}^{(L,m)}}{\epsilon^m}
 \ee
where $L$ is the loop order.
 $\gamma_{\Phi}$ is determined by the simple poles in $Z_{\Phi}$ according
 to
 \be
 \gamma_{\Phi}^{(L)}=-\frac{L}{2}Z_{\Phi}^{(L,1)}.
\label{gamdef} 
\ee
The anomalous dimension of the chiral superfield is given at two loops
 by\cite{penatib,ASW}
 \be
 (8\pi)^2\gamma_{\Phi}^{(2)}=
 \tfrac13Y_2-\tfrac{2}{k^2}C_RC_R-\tfrac{1}{k^2}\Ttil C_R
 \label{gamphi}
 \ee
 where
 \begin{align}
 (Y_2)_i{}^j=&\Ybar_{iklm}Y^{jklm}\nn
 C_R=&R_AR_A,\nn
 C_G\delta_{AB}=&f_{ACD}f_{BCD},\nn
\Ttil&=T_R-C_G,
\label{defs}
 \end{align}
 and $T_R$ is defined in Eq.~(\ref{Tdef}).
 This result may readily be obtained by $\Ncal=2$ superfield
 methods\cite{kaza,GN,ASW,penatib}; see
 Appendix A for our $\Ncal=2$ superfield conventions. Henceforth we set
 $k=1$ for simplicity, and also neglect factors of $(8\pi)^2$ (one for each loop order); these factors of $k$ and $(8\pi)^2$ may easily be restored if desired. The two-loop
 results for general Chern-Simons theories  obtained in
 Ref.~\cite{kazb} are not directly comparable since they were computed
 in the $\Ncal=1$ framework.

The $\beta$-functions $\beta_Y$ and $\beta_{\Ybar}$ are given at lowest order by  inserting \eqref{gamphi} into \eqref{nonren}. It is then clear that Eq.~\eqref{grad} is satisfied to this order in the form\footnote{We prefer to use the notation $A$ (rather than $a$) in three dimensions in the absence of any connection to a conformal anomaly coefficient.}
\be
\frac{\pa}{\pa Y^{ijkl}}A=\beta_{\Ybar ijkl},\quad \frac{\pa}{\pa \Ybar_{ijkl} }A=\beta_Y^{ ijkl},
\label{gradtwo}
\ee
(hence effectively with a unit $T_{IJ}$ on the right-hand side of Eq.~\eqref{grad}) by taking
\be
A^{(5)}=a^{(5)}_1A_1^{(5)}+a^{(5)}_2A_2^{(5)}+a^{(5)}_3A_3^{(5)}
\ee
where
\be
A_1^{(5)}=(Y_2)_i{}^j(Y_2)_j{}^i,\quad A_2^{(5)}=(Y_2)_i{}^j(C_R^2)_j{}^i,\quad
 A_3^{(5)}=C_G(Y_2)_i{}^j(C_R)_j{}^i,
\ee
with $Y_2$ given by Eq.~\eqref{defs}, and
\be
a^{(5)}_1=\tfrac23,\quad a^{(5)}_2=-8,\quad a^{(5)}_3=-4\Ttil.
\ee
Notice that at this order the construction of the $a$-function imposes no constraints on the anomalous dimension coefficients, since there is a one-to-one correspondence between $a$-function structures and anomalous dimension
structures. For later convenience it will be useful to rewrite Eq.~\eqref{gradtwo} in the form
\be
d_YA\equiv dY\circ\pa_Y A=dY\circ \beta_{\Ybar},\quad d_{\Ybar}A\equiv d\Ybar\circ\pa_{\Ybar} A=d\Ybar\circ \beta_Y,
\ee 
where $\circ$ is a scalar product on Yukawa couplings so that for instance $Y\circ \Ybar\equiv Y^{ijkl}\Ybar_{ijkl}$.

 \section{Consistency conditions for four-loop anomalous dimension}
In this section we derive the consistency conditions on the four-loop anomalous dimension coefficients required for Eq.~\eqref{grad} to be satisfied at next-to-leading order. We assume an $a$-function at this order of the general form
\be
A^{(7)}=\sum_{i=1}^{14}a^{(7)}_iA_i^{(7)}+\ta(\beta_Y^{(2)})\circ (\beta_{\Ybar}^{(2)})
\label{adef}
\ee

\noindent where the structures $ A_i^{(7)} $ are depicted in Table~\ref{fig1}, except for $A_{14}^{(7)}$ which it is more convenient simply to define explicitly, {\it viz}:
\be
A_{14}^{(7)}=\tfrac14\tr(Y_2\{R_A,R_B\})\tr(Y_2\{R_A,R_B\}),
\ee
 and the final term represents the usual arbitrariness\cite{OsbJacnew} in defining the $a$-function. Our convention for the chiral lines is that the arrows point from a $Y$ vertex to a $\Ybar$ vertex; furthermore, a box represents an insertion of $C_R$, and an $A$ or $B$ represents an insertion of a gauge generator $R_A$ or $R_B$ respectively. At this order we expect 
\be
d_YA^{(7)}=dY\circ T^{(3)}\circ\beta_{\Ybar}^{(4)}+dY\circ T^{(5)}\circ \beta_{\Ybar}^{(2)}+dY\circ K^{(5)}\circ \beta_{Y}^{(2)}
\ee
where as we saw in the previous section, $T^{(3)}$ is effectively the unit matrix and we write
\be
T^{(5)}=\sum_{i=1}^{7}t^{(5)}_iT_i^{(5)},\quad K^{(5)}=\sum_{i=1}^{2}k^{(5)}_iK_i^{(5)},
\ee
where the individual contributions $dY\circ T_i^{(5)}\circ d\Ybar$, $dY\circ K_i^{(5)}\circ dY$ are depicted in Tables~\ref{fig2} and \ref{fig3}, with a cross denoting $dY$ and a diamond denoting $d\Ybar$. The corresponding expression for $d_{\Ybar}$ may be obtained by conjugation and is not given explicitly. Finally the four-loop
 anomalous dimension is expected to take the form
 \be
 (8\pi)^4\gamma_{\Phi}^{(4)}=\sum_{i=1}^{12}g_i\gamma^{(4)}_i+\ldots
\label{gdef}
 \ee
 where the invariants involving Yukawa couplings are given by
 
 \begin{table}[t]
 	\setlength{\extrarowheight}{0.5cm}
 	\setlength{\tabcolsep}{24pt}
 	\hspace*{-7.25cm}
 	\centering
 	\resizebox{6.5cm}{!}{
 		\begin{tabular*}{20cm}{ccccc}
 			\begin{picture}(182,166) (309,-207)
 			\SetWidth{1.0}
 			\SetColor{Black}
 			\Arc[arrow,arrowpos=0.13,arrowlength=15,arrowwidth=10,arrowinset=1](400,-122)(80.056,131,491)
 			\SetWidth{0.0}
 			\Arc[arrow,arrowpos=0.3,arrowlength=15,arrowwidth=10,arrowinset=1,flip](398.333,-120.333)(78.351,88.781,181.219)
 			\Arc[arrow,arrowpos=0.3,arrowlength=15,arrowwidth=10,arrowinset=1,flip,clock](401.667,-120.333)(78.351,91.219,-1.219)
 			\Arc[arrow,arrowpos=0.7,arrowlength=15,arrowwidth=10,arrowinset=1,flip](398.333,-123.667)(78.351,178.781,271.219)
 			\Arc[arrow,arrowpos=0.7,arrowlength=15,arrowwidth=10,arrowinset=1,flip,clock](401.667,-123.667)(78.351,1.219,-91.219)
 			\Arc[arrow,arrowpos=0.5,arrowlength=15,arrowwidth=10,arrowinset=1,clock](400,-122)(80,90,-90)
 			\SetWidth{1.0}
 			\Arc[arrow,arrowpos=0.5,arrowlength=15,arrowwidth=10,arrowinset=1,clock](311.333,-122)(56.667,61.928,-61.928)
 			\Arc[arrow,arrowpos=0.5,arrowlength=15,arrowwidth=10,arrowinset=1](488.667,-122)(56.667,118.072,241.928)
 			\Arc[arrow,arrowpos=0.5,arrowlength=15,arrowwidth=10,arrowinset=1,clock](132.667,-122)(211.333,13.686,-13.686)
 			\Arc[arrow,arrowpos=0.5,arrowlength=15,arrowwidth=10,arrowinset=1](667.333,-122)(211.333,166.314,193.686)
 			\Arc[arrow,arrowpos=0.5,arrowlength=15,arrowwidth=10,arrowinset=1,flip](660.667,-122)(272.667,162.938,197.062)
 			\Arc[arrow,arrowpos=0.5,arrowlength=15,arrowwidth=10,arrowinset=1,flip,clock](139.333,-122)(272.667,17.062,-17.062)
 			\end{picture}
 			&
 			\begin{picture}(182,166) (309,-207)
 			\SetWidth{1.0}
 			\SetColor{Black}
 			\Arc[arrow,arrowpos=0.13,arrowlength=15,arrowwidth=10,arrowinset=1](400,-122)(80.056,131,491)
 			\SetWidth{0.0}
 			\Arc[arrow,arrowpos=0.3,arrowlength=15,arrowwidth=10,arrowinset=1,flip](398.333,-120.333)(78.351,88.781,181.219)
 			\Arc[arrow,arrowpos=0.5,arrowlength=15,arrowwidth=10,arrowinset=1,flip,clock](401.986,-116.963)(74.989,91.517,36.841)
 			\Arc[arrow,arrowpos=0.7,arrowlength=15,arrowwidth=10,arrowinset=1,flip](398.333,-123.667)(78.351,178.781,271.219)
 			\Arc[arrow,arrowpos=0.7,arrowlength=15,arrowwidth=10,arrowinset=1,flip,clock](401.667,-123.667)(78.351,1.219,-91.219)
 			\Arc[arrow,arrowpos=0.5,arrowlength=15,arrowwidth=10,arrowinset=1,clock](400,-122)(80,90,-90)
 			\SetWidth{1.0}
 			\Arc[arrow,arrowpos=0.5,arrowlength=15,arrowwidth=10,arrowinset=1,clock](311.333,-122)(56.667,61.928,-61.928)
 			\Arc[arrow,arrowpos=0.5,arrowlength=15,arrowwidth=10,arrowinset=1,clock](132.667,-122)(211.333,13.686,-13.686)
 			\Arc[arrow,arrowpos=0.5,arrowlength=15,arrowwidth=10,arrowinset=1](560.167,-122)(110.167,153.009,206.991)
 			\Line[arrow,arrowpos=0.5,arrowlength=15,arrowwidth=10,arrowinset=1,flip](400,-42)(400,-202)
 			\Arc[arrow,arrowpos=0.5,arrowlength=15,arrowwidth=10,arrowinset=1,flip](441.469,-35.364)(41.997,-170.908,-60.734)
 			\Arc[arrow,arrowpos=0.5,arrowlength=15,arrowwidth=10,arrowinset=1,clock](441.469,-208.636)(41.997,170.908,60.734)
 			\end{picture}
 			&
 			\begin{picture}(178,182) (311,-197)
 			\SetWidth{1.0}
 			\SetColor{Black}
 			\Arc[arrow,arrowpos=0.875,arrowlength=15,arrowwidth=10,arrowinset=1,flip](400,-106)(80.056,131,491)
 			\Arc[arrow,arrowpos=0.5,arrowlength=15,arrowwidth=10,arrowinset=1,clock](400,-217.261)(77.261,135.388,44.612)
 			\Line[arrow,arrowpos=0.5,arrowlength=15,arrowwidth=10,arrowinset=1](345,-163)(455,-163)
 			\SetWidth{0.0}
 			\Arc[arrow,arrowpos=0.5,arrowlength=15,arrowwidth=10,arrowinset=1](400,-108.739)(77.261,-135.388,-44.612)
 			\SetWidth{1.0}
 			\Arc[arrow,arrowpos=0.5,arrowlength=15,arrowwidth=10,arrowinset=1,clock](304.104,-57.073)(73.665,20.729,-74.33)
 			\Line[arrow,arrowpos=0.5,arrowlength=15,arrowwidth=10,arrowinset=1](373,-31)(324,-128)
 			\SetWidth{0.0}
 			\Arc[arrow,arrowpos=0.5,arrowlength=15,arrowwidth=10,arrowinset=1](399.572,-105.299)(78.908,109.679,196.719)
 			\SetWidth{1.0}
 			\Arc[arrow,arrowpos=0.5,arrowlength=15,arrowwidth=10,arrowinset=1,flip](495.896,-57.073)(73.665,159.271,254.33)
 			\Line[arrow,arrowpos=0.5,arrowlength=15,arrowwidth=10,arrowinset=1,flip](427,-31)(476,-128)
 			\SetWidth{0.0}
 			\Arc[arrow,arrowpos=0.5,arrowlength=15,arrowwidth=10,arrowinset=1,flip,clock](400.428,-105.299)(78.908,70.321,-16.719)
 			\Arc[arrow,arrowpos=0.5,arrowlength=15,arrowwidth=10,arrowinset=1,clock](411.368,-112.053)(67.077,-15.519,-49.423)
 			\Arc[arrow,arrowpos=0.5,arrowlength=15,arrowwidth=10,arrowinset=1,clock](381.405,-117.357)(58.383,-128.576,-169.497)
 			\end{picture}
 			&
 			\begin{picture}(182,183) (309,-197)
 			\SetWidth{1.0}
 			\SetColor{Black}
 			\Arc[arrow,arrowpos=0.875,arrowlength=15,arrowwidth=10,arrowinset=1,flip](400,-105)(80.056,131,491)
 			\Arc[arrow,arrowpos=0.5,arrowlength=15,arrowwidth=10,arrowinset=1,clock](305,-105)(80,53.13,-53.13)
 			\Line[arrow,arrowpos=0.5,arrowlength=15,arrowwidth=10,arrowinset=1](353,-41)(353,-169)
 			\SetWidth{0.0}
 			\Arc[arrow,arrowpos=0.5,arrowlength=15,arrowwidth=10,arrowinset=1](398.561,-105)(78.561,125.446,234.554)
 			\SetWidth{1.0}
 			\Arc[arrow,arrowpos=0.5,arrowlength=15,arrowwidth=10,arrowinset=1,flip](495,-105)(80,126.87,233.13)
 			\SetWidth{0.0}
 			\Arc[arrow,arrowpos=0.35,arrowlength=15,arrowwidth=10,arrowinset=1,flip](400,-105)(80,-180,0)
 			\SetWidth{1.0}
 			\Line[arrow,arrowpos=0.5,arrowlength=15,arrowwidth=10,arrowinset=1,flip](447,-41)(447,-169)
 			\SetWidth{0.0}
 			\CBox(397.314,-196.314)(374.686,-173.686){Black}{Black}
 			\Arc[arrow,arrowpos=0.87,arrowlength=15,arrowwidth=10,arrowinset=1,flip](400,-107.969)(77.031,-127.6,-52.4)
 			\CBox(425.314,-196.314)(402.686,-173.686){Black}{Black}
 			\Arc[arrow,arrowpos=0.5,arrowlength=15,arrowwidth=10,arrowinset=1](400,-105)(80,-90,90)
 			\end{picture}
 			&
 			\begin{picture}(180,182) (309,-197)
 			\SetWidth{1.0}
 			\SetColor{Black}
 			\Arc[arrow,arrowpos=0.875,arrowlength=15,arrowwidth=10,arrowinset=1,flip](400,-106)(80.056,131,491)
 			\Arc[arrow,arrowpos=0.5,arrowlength=15,arrowwidth=10,arrowinset=1,clock](305,-106)(80,53.13,-53.13)
 			\Line[arrow,arrowpos=0.5,arrowlength=15,arrowwidth=10,arrowinset=1](353,-42)(353,-170)
 			\SetWidth{0.0}
 			\Arc[arrow,arrowpos=0.5,arrowlength=15,arrowwidth=10,arrowinset=1](398.561,-106)(78.561,125.446,234.554)
 			\SetWidth{1.0}
 			\Arc[arrow,arrowpos=0.5,arrowlength=15,arrowwidth=10,arrowinset=1,flip](495,-106)(80,126.87,233.13)
 			\Line[arrow,arrowpos=0.5,arrowlength=15,arrowwidth=10,arrowinset=1,flip](447,-42)(447,-170)
 			\SetWidth{0.0}
 			\Arc[arrow,arrowpos=0.35,arrowlength=15,arrowwidth=10,arrowinset=1,flip,clock](402.679,-105.361)(77.324,55.027,-0.474)
 			\Arc[arrow,arrowpos=0.65,arrowlength=15,arrowwidth=10,arrowinset=1,flip,clock](396.407,-103.405)(83.633,-1.778,-52.776)
 			\SetWidth{1.0}
 			\CBox(487.662,-99.662)(464.338,-76.338){Black}{Black}
 			\CBox(487.662,-135.662)(464.338,-112.338){Black}{Black}
 			\SetWidth{0.0}
 			\Arc[arrow,arrowpos=0.5,arrowlength=15,arrowwidth=10,arrowinset=1,flip](400,-106)(80,-180,0)
 			\end{picture}
 			\\
 			{\Huge $A^{(7)}_{1}$}
 			&
 			{\Huge $A^{(7)}_{2}$}
 			&
 			{\Huge $A^{(7)}_{3}$}
 			&	
 			{\Huge $A^{(7)}_{4}$}
 			&
 			{\Huge $A^{(7)}_{5}$}
 			\\
 			&
 			&
 			&
 			&
 			\\
 			\begin{picture}(184,184) (309,-197)
 			\SetWidth{1.0}
 			\SetColor{Black}
 			\Arc[arrow,arrowpos=0.875,arrowlength=15,arrowwidth=10,arrowinset=1,flip](400,-104)(80.056,131,491)
 			\Arc[arrow,arrowpos=0.5,arrowlength=15,arrowwidth=10,arrowinset=1,clock](305,-104)(80,53.13,-53.13)
 			\Line[arrow,arrowpos=0.5,arrowlength=15,arrowwidth=10,arrowinset=1](353,-40)(353,-168)
 			\SetWidth{0.0}
 			\Arc[arrow,arrowpos=0.5,arrowlength=15,arrowwidth=10,arrowinset=1](398.561,-104)(78.561,125.446,234.554)
 			\SetWidth{1.0}
 			\Arc[arrow,arrowpos=0.5,arrowlength=15,arrowwidth=10,arrowinset=1,flip](495,-104)(80,126.87,233.13)
 			\SetWidth{0.0}
 			\Arc[arrow,arrowpos=0.55,arrowlength=15,arrowwidth=10,arrowinset=1,flip,clock](402.679,-103.361)(77.324,55.027,-0.474)
 			\Arc[arrow,arrowpos=0.45,arrowlength=15,arrowwidth=10,arrowinset=1,flip,clock](396.407,-101.405)(83.633,-1.778,-52.776)
 			\Arc[arrow,arrowpos=0.37,arrowlength=15,arrowwidth=10,arrowinset=1,flip](400,-104)(80,-180,0)
 			\SetWidth{1.0}
 			\Line[arrow,arrowpos=0.5,arrowlength=15,arrowwidth=10,arrowinset=1,flip](447,-40)(447,-168)
 			\CBox(491.662,-115.662)(468.338,-92.338){Black}{Black}
 			\CBox(412.083,-196.083)(387.917,-171.917){Black}{Black}
 			\SetWidth{0.0}
 			\Arc[arrow,arrowpos=0.78,arrowlength=15,arrowwidth=10,arrowinset=1,flip](400,-106.969)(77.031,-127.6,-52.4)
 			\end{picture}
 			&
 			\begin{picture}(184,182) (309,-197)
 			\SetWidth{1.0}
 			\SetColor{Black}
 			\Arc[arrow,arrowpos=0.875,arrowlength=15,arrowwidth=10,arrowinset=1,flip](400,-106)(80.056,131,491)
 			\Arc[arrow,arrowpos=0.5,arrowlength=15,arrowwidth=10,arrowinset=1,clock](305,-106)(80,53.13,-53.13)
 			\Line[arrow,arrowpos=0.5,arrowlength=15,arrowwidth=10,arrowinset=1](353,-42)(353,-170)
 			\SetWidth{0.0}
 			\Arc[arrow,arrowpos=0.5,arrowlength=15,arrowwidth=10,arrowinset=1](398.561,-106)(78.561,125.446,234.554)
 			\SetWidth{1.0}
 			\Arc[arrow,arrowpos=0.5,arrowlength=15,arrowwidth=10,arrowinset=1,flip](495,-106)(80,126.87,233.13)
 			\SetWidth{0.0}
 			\Arc[arrow,arrowpos=0.55,arrowlength=15,arrowwidth=10,arrowinset=1,flip,clock](402.679,-105.361)(77.324,55.027,-0.474)
 			\Arc[arrow,arrowpos=0.45,arrowlength=15,arrowwidth=10,arrowinset=1,flip,clock](396.407,-103.405)(83.633,-1.778,-52.776)
 			\SetWidth{1.0}
 			\CBox(491.662,-117.662)(468.338,-94.338){Black}{Black}
 			\SetWidth{0.0}
 			\Arc[arrow,arrowpos=0.5,arrowlength=15,arrowwidth=10,arrowinset=1,flip](400,-106)(80,-180,0)
 			\SetWidth{1.0}
 			\Line[arrow,arrowpos=0.5,arrowlength=15,arrowwidth=10,arrowinset=1,flip](447,-42)(447,-106)
 			\Line[arrow,arrowpos=0.5,arrowlength=15,arrowwidth=10,arrowinset=1,flip](447,-106)(447,-170)
 			\CBox(458.662,-117.662)(435.338,-94.338){Black}{Black}
 			\end{picture}
 			&
 			\begin{picture}(203,193) (309,-186)
 			\SetWidth{1.0}
 			\SetColor{Black}
 			\Arc[arrow,arrowpos=0.95,arrowlength=15,arrowwidth=10,arrowinset=1,flip](400,-95)(80.056,131,491)
 			\Arc[arrow,arrowpos=0.5,arrowlength=15,arrowwidth=10,arrowinset=1,clock](305,-95)(80,53.13,-53.13)
 			\Line[arrow,arrowpos=0.5,arrowlength=15,arrowwidth=10,arrowinset=1](353,-31)(353,-159)
 			\SetWidth{0.0}
 			\Arc[arrow,arrowpos=0.5,arrowlength=15,arrowwidth=10,arrowinset=1](398.561,-95)(78.561,125.446,234.554)
 			\SetWidth{1.0}
 			\Arc[arrow,arrowpos=0.5,arrowlength=15,arrowwidth=10,arrowinset=1,flip](495,-95)(80,126.87,233.13)
 			\SetWidth{0.0}
 			\Arc[arrow,arrowpos=0.55,arrowlength=15,arrowwidth=10,arrowinset=1,flip,clock](402.679,-94.361)(77.324,55.027,-0.474)
 			\Arc[arrow,arrowpos=0.45,arrowlength=15,arrowwidth=10,arrowinset=1,flip,clock](396.407,-92.405)(83.633,-1.778,-52.776)
 			\Arc[arrow,arrowpos=0.5,arrowlength=15,arrowwidth=10,arrowinset=1,flip](400,-95)(80,-180,0)
 			\Text(495,-83)[]{\Huge{\Black{$B$}}}
 			\Text(495,-107)[]{\Huge{\Black{$A$}}}
 			\SetWidth{1.0}
 			\Line[arrow,arrowpos=0.5,arrowlength=15,arrowwidth=10,arrowinset=1,flip](447,-31)(447,-159)
 			\Text(383,-3)[]{\Huge{\Black{$A$}}}
 			\Text(417,-3)[]{\Huge{\Black{$B$}}}
 			\Arc[arrow,arrowpos=0.65,arrowlength=15,arrowwidth=10,arrowinset=1,clock](400,-95)(80,-180,-360)
 			\end{picture}
 			&
 			\begin{picture}(203,182) (309,-197)
 			\SetWidth{1.0}
 			\SetColor{Black}
 			\Arc[arrow,arrowpos=0.875,arrowlength=15,arrowwidth=10,arrowinset=1,flip](400,-106)(80.056,131,491)
 			\Arc[arrow,arrowpos=0.5,arrowlength=15,arrowwidth=10,arrowinset=1,clock](305,-106)(80,53.13,-53.13)
 			\Line[arrow,arrowpos=0.5,arrowlength=15,arrowwidth=10,arrowinset=1](353,-42)(353,-170)
 			\SetWidth{0.0}
 			\Arc[arrow,arrowpos=0.5,arrowlength=15,arrowwidth=10,arrowinset=1](398.561,-106)(78.561,125.446,234.554)
 			\SetWidth{1.0}
 			\Arc[arrow,arrowpos=0.5,arrowlength=15,arrowwidth=10,arrowinset=1,flip](495,-106)(80,126.87,233.13)
 			\SetWidth{0.0}
 			\Arc[arrow,arrowpos=0.55,arrowlength=15,arrowwidth=10,arrowinset=1,flip,clock](402.679,-105.361)(77.324,55.027,-0.474)
 			\Arc[arrow,arrowpos=0.45,arrowlength=15,arrowwidth=10,arrowinset=1,flip,clock](396.407,-103.405)(83.633,-1.778,-52.776)
 			\Arc[arrow,arrowpos=0.5,arrowlength=15,arrowwidth=10,arrowinset=1,flip](400,-106)(80,-180,0)
 			\SetWidth{1.0}
 			\Line[arrow,arrowpos=0.5,arrowlength=15,arrowwidth=10,arrowinset=1,flip](447,-42)(447,-106)
 			\Line[arrow,arrowpos=0.5,arrowlength=15,arrowwidth=10,arrowinset=1,flip](447,-106)(447,-170)
 			\Text(460,-94)[]{\Huge{\Black{$A$}}}
 			\Text(460,-118)[]{\Huge{\Black{$B$}}}
 			\Text(495,-94)[]{\Huge{\Black{$A$}}}
 			\Text(495,-118)[]{\Huge{\Black{$B$}}}
 			\end{picture}
 			&
 			\begin{picture}(171,167) (319,-207)
 			\SetWidth{1.0}
 			\SetColor{Black}
 			\Arc[arrow,arrowpos=0,arrowlength=15,arrowwidth=10,arrowinset=1,flip](400,-121)(79.981,132,492)
 			\Arc[arrow,arrowpos=0.5,arrowlength=15,arrowwidth=10,arrowinset=1,flip,clock](321.441,-42.441)(78.572,1.051,-91.051)
 			\Arc[arrow,arrowpos=0.5,arrowlength=15,arrowwidth=10,arrowinset=1,flip](478.559,-42.441)(78.572,178.949,271.051)
 			\Arc[arrow,arrowpos=0.5,arrowlength=15,arrowwidth=10,arrowinset=1,flip](321.441,-199.559)(78.572,-1.051,91.051)
 			\Arc[arrow,arrowpos=0.5,arrowlength=15,arrowwidth=10,arrowinset=1,flip,clock](478.559,-199.559)(78.572,-178.949,-271.051)
 			\SetWidth{0.0}
 			\Arc[arrow,arrowpos=0.5,arrowlength=15,arrowwidth=10,arrowinset=1,flip,clock](401.441,-119.559)(78.572,91.051,-1.051)
 			\Arc[arrow,arrowpos=0.175,arrowlength=15,arrowwidth=10,arrowinset=1,flip](401.441,-122.441)(78.572,-91.051,1.051)
 			\Arc[arrow,arrowpos=0.5,arrowlength=15,arrowwidth=10,arrowinset=1,flip,clock](398.559,-122.441)(78.572,-88.949,-181.051)
 			\Arc[arrow,arrowpos=0.85,arrowlength=15,arrowwidth=10,arrowinset=1,flip](401.441,-122.441)(78.572,-91.051,1.051)
 			\SetWidth{1.0}
 			\CBox(480.042,-175.042)(455.958,-150.958){Black}{Black}
 			\CBox(455.042,-201.042)(430.958,-176.958){Black}{Black}
 			\end{picture}
 			\\
 			{\Huge $A^{(7)}_{6}$}
 			&
 			{\Huge $A^{(7)}_{7}$}
 			&	
 			{\Huge $A^{(7)}_{8}$}
 			&
 			{\Huge $A^{(7)}_{9}$}
 			&
 			{\Huge $A^{(7)}_{10}$}
 			\\
 			&
 			&
 			&
 			&
 			\\
 			\begin{picture}(168,163) (319,-207)
 			\SetWidth{1.0}
 			\SetColor{Black}
 			\Arc[arrow,arrowpos=0,arrowlength=15,arrowwidth=10,arrowinset=1,flip](400,-125)(79.981,132,492)
 			\Arc[arrow,arrowpos=0.5,arrowlength=15,arrowwidth=10,arrowinset=1,flip,clock](321.441,-46.441)(78.572,1.051,-91.051)
 			\Arc[arrow,arrowpos=0.5,arrowlength=15,arrowwidth=10,arrowinset=1,flip](478.559,-46.441)(78.572,178.949,271.051)
 			\Arc[arrow,arrowpos=0.5,arrowlength=15,arrowwidth=10,arrowinset=1,flip](321.441,-203.559)(78.572,-1.051,91.051)
 			\Arc[arrow,arrowpos=0.5,arrowlength=15,arrowwidth=10,arrowinset=1,flip,clock](478.559,-203.559)(78.572,-178.949,-271.051)
 			\SetWidth{0.0}
 			\Arc[arrow,arrowpos=0.5,arrowlength=15,arrowwidth=10,arrowinset=1,flip,clock](401.441,-123.559)(78.572,91.051,-1.051)
 			\Arc[arrow,arrowpos=0.25,arrowlength=15,arrowwidth=10,arrowinset=1,flip](401.441,-126.441)(78.572,-91.051,1.051)
 			\Arc[arrow,arrowpos=0.5,arrowlength=15,arrowwidth=10,arrowinset=1,flip,clock](398.559,-126.441)(78.572,-88.949,-181.051)
 			\Arc[arrow,arrowpos=0.75,arrowlength=15,arrowwidth=10,arrowinset=1,flip](401.441,-126.441)(78.572,-91.051,1.051)
 			\SetWidth{1.0}
 			\CBox(469.042,-194.042)(444.958,-169.958){Black}{Black}
 			\end{picture}
 			&
 			\begin{picture}(182,184) (309,-197)
 			\SetWidth{1.0}
 			\SetColor{Black}
 			\Arc[arrow,arrowpos=0.875,arrowlength=15,arrowwidth=10,arrowinset=1,flip](400,-104)(80.056,131,491)
 			\Arc[arrow,arrowpos=0.5,arrowlength=15,arrowwidth=10,arrowinset=1,clock](305,-104)(80,53.13,-53.13)
 			\Line[arrow,arrowpos=0.5,arrowlength=15,arrowwidth=10,arrowinset=1](353,-40)(353,-168)
 			\SetWidth{0.0}
 			\Arc[arrow,arrowpos=0.5,arrowlength=15,arrowwidth=10,arrowinset=1](398.561,-104)(78.561,125.446,234.554)
 			\SetWidth{1.0}
 			\Arc[arrow,arrowpos=0.5,arrowlength=15,arrowwidth=10,arrowinset=1,flip](495,-104)(80,126.87,233.13)
 			\SetWidth{0.0}
 			\Arc[arrow,arrowpos=0.37,arrowlength=15,arrowwidth=10,arrowinset=1,flip](400,-104)(80,-180,0)
 			\SetWidth{1.0}
 			\Line[arrow,arrowpos=0.5,arrowlength=15,arrowwidth=10,arrowinset=1,flip](447,-40)(447,-168)
 			\CBox(412.083,-196.083)(387.917,-171.917){Black}{Black}
 			\SetWidth{0.0}
 			\Arc[arrow,arrowpos=0.78,arrowlength=15,arrowwidth=10,arrowinset=1,flip](400,-106.969)(77.031,-127.6,-52.4)
 			\Arc[arrow,arrowpos=0.5,arrowlength=15,arrowwidth=10,arrowinset=1,flip,clock](400,-104)(80,90,-90)
 			\end{picture}
 			&
 			\begin{picture}(184,182) (309,-197)
 			\SetWidth{1.0}
 			\SetColor{Black}
 			\Arc[arrow,arrowpos=0.875,arrowlength=15,arrowwidth=10,arrowinset=1,flip](400,-106)(80.056,131,491)
 			\Arc[arrow,arrowpos=0.5,arrowlength=15,arrowwidth=10,arrowinset=1,clock](305,-106)(80,53.13,-53.13)
 			\Line[arrow,arrowpos=0.5,arrowlength=15,arrowwidth=10,arrowinset=1](353,-42)(353,-170)
 			\SetWidth{0.0}
 			\Arc[arrow,arrowpos=0.5,arrowlength=15,arrowwidth=10,arrowinset=1](398.561,-106)(78.561,125.446,234.554)
 			\SetWidth{1.0}
 			\Arc[arrow,arrowpos=0.5,arrowlength=15,arrowwidth=10,arrowinset=1,flip](495,-106)(80,126.87,233.13)
 			\Line[arrow,arrowpos=0.5,arrowlength=15,arrowwidth=10,arrowinset=1,flip](447,-42)(447,-170)
 			\SetWidth{0.0}
 			\Arc[arrow,arrowpos=0.35,arrowlength=15,arrowwidth=10,arrowinset=1,flip,clock](402.679,-105.361)(77.324,55.027,-0.474)
 			\Arc[arrow,arrowpos=0.65,arrowlength=15,arrowwidth=10,arrowinset=1,flip,clock](396.407,-103.405)(83.633,-1.778,-52.776)
 			\SetWidth{1.0}
 			\CBox(491.662,-117.662)(468.338,-94.338){Black}{Black}
 			\SetWidth{0.0}
 			\Arc[arrow,arrowpos=0.5,arrowlength=15,arrowwidth=10,arrowinset=1,flip](400,-106)(80,-180,0)
 			\end{picture}
 			&
 			
 			&
 			\\
 			{\Huge $A^{(7)}_{11}$}
 			&	
 			{\Huge $A^{(7)}_{12}$}
 			&
 			{\Huge $A^{(7)}_{13}$}
 			&
 			
 			&
 		\end{tabular*}
 	}
 	\caption{Contributions to $A^{(7)}$ (for notation, see Section 3)}
 	\label{fig1}
 \end{table}
 
 \begin{table}[t]
 	\setlength{\extrarowheight}{0.5cm}
 	\setlength{\tabcolsep}{24pt}
 	\hspace*{-7.25cm}
 	\centering
 	\resizebox{6.5cm}{!}{
 		\begin{tabular*}{20cm}{cccccc}
 			\begin{picture}(181,182) (310,-197)
 			\SetWidth{1.0}
 			\SetColor{Black}
 			\CTri(345.447,-176)(362,-159.447)(378.553,-176){Black}{Black}\CTri(345.447,-176)(362,-192.553)(378.553,-176){Black}{Black}
 			\Arc[arrow,arrowpos=0.5,arrowlength=15,arrowwidth=10,arrowinset=1,flip](400,-106)(80,-180,0)
 			\SetWidth{1.5}
 			\Line(347,-21)(377,-51)
 			\Line(377,-21)(347,-51)
 			\SetWidth{1.0}
 			\Arc[arrow,arrowpos=0.5,arrowlength=15,arrowwidth=10,arrowinset=1,clock](400,-106)(80,-180,-360)
 			\Arc[arrow,arrowpos=0.125,arrowlength=15,arrowwidth=10,arrowinset=1](400,-106)(79.981,132,492)
 			\Arc[arrow,arrowpos=0.48,arrowlength=15,arrowwidth=10,arrowinset=1,clock](261.636,-106)(122.364,34.894,-34.894)
 			\Arc[arrow,arrowpos=0.48,arrowlength=15,arrowwidth=10,arrowinset=1,flip](538.364,-106)(122.364,145.106,214.894)
 			\Arc[arrow,arrowpos=0.48,arrowlength=15,arrowwidth=10,arrowinset=1](602,-106)(250,163.74,196.26)
 			\Arc[arrow,arrowpos=0.48,arrowlength=15,arrowwidth=10,arrowinset=1,flip,clock](198,-106)(250,16.26,-16.26)
 			\SetWidth{0.0}
 			\Arc[arrow,arrowpos=0.48,arrowlength=15,arrowwidth=10,arrowinset=1,flip,clock](400.667,-106)(79.333,61.928,-61.928)
 			\end{picture}
 			&
 			\begin{picture}(181,182) (310,-197)
 			\SetWidth{1.0}
 			\SetColor{Black}
 			\CTri(345.447,-176)(362,-159.447)(378.553,-176){Black}{Black}\CTri(345.447,-176)(362,-192.553)(378.553,-176){Black}{Black}
 			\Arc[arrow,arrowpos=0.5,arrowlength=15,arrowwidth=10,arrowinset=1,flip](400,-106)(80,-180,0)
 			\SetWidth{1.5}
 			\Line(423,-161)(453,-191)
 			\Line(453,-161)(423,-191)
 			\SetWidth{1.0}
 			\Arc[arrow,arrowpos=0.5,arrowlength=15,arrowwidth=10,arrowinset=1,clock](400,-106)(80,-180,-360)
 			\Arc[arrow,arrowpos=0.125,arrowlength=15,arrowwidth=10,arrowinset=1](400,-106)(79.981,132,492)
 			\Arc[arrow,arrowpos=0.48,arrowlength=15,arrowwidth=10,arrowinset=1,clock](261.636,-106)(122.364,34.894,-34.894)
 			\Arc[arrow,arrowpos=0.48,arrowlength=15,arrowwidth=10,arrowinset=1,flip](538.364,-106)(122.364,145.106,214.894)
 			\Arc[arrow,arrowpos=0.48,arrowlength=15,arrowwidth=10,arrowinset=1](602,-106)(250,163.74,196.26)
 			\Arc[arrow,arrowpos=0.48,arrowlength=15,arrowwidth=10,arrowinset=1,flip,clock](198,-106)(250,16.26,-16.26)
 			\SetWidth{0.0}
 			\Arc[arrow,arrowpos=0.48,arrowlength=15,arrowwidth=10,arrowinset=1,flip,clock](400.667,-106)(79.333,61.928,-61.928)
 			\end{picture}
 			&
 			\begin{picture}(178,179) (303,-207)
 			\SetWidth{1.0}
 			\SetColor{Black}
 			\CTri(383.447,-189)(400,-172.447)(416.553,-189){Black}{Black}\CTri(383.447,-189)(400,-205.553)(416.553,-189){Black}{Black}
 			\SetWidth{1.5}
 			\Line(305,-94)(335,-124)
 			\Line(335,-94)(305,-124)
 			\SetWidth{1.0}
 			\Arc[arrow,arrowpos=0,arrowlength=15,arrowwidth=10,arrowinset=1,flip](400,-109)(79.981,132,492)
 			\Arc[arrow,arrowpos=0.5,arrowlength=15,arrowwidth=10,arrowinset=1,flip,clock](321.441,-30.441)(78.572,1.051,-91.051)
 			\Arc[arrow,arrowpos=0.5,arrowlength=15,arrowwidth=10,arrowinset=1,flip](478.559,-30.441)(78.572,178.949,271.051)
 			\Arc[arrow,arrowpos=0.5,arrowlength=15,arrowwidth=10,arrowinset=1,flip](321.441,-187.559)(78.572,-1.051,91.051)
 			\Arc[arrow,arrowpos=0.5,arrowlength=15,arrowwidth=10,arrowinset=1,flip,clock](478.559,-187.559)(78.572,-178.949,-271.051)
 			\SetWidth{0.0}
 			\Arc[arrow,arrowpos=0.5,arrowlength=15,arrowwidth=10,arrowinset=1,flip,clock](400.379,-108.621)(79.622,90.273,-0.273)
 			\Arc[arrow,arrowpos=0.5,arrowlength=15,arrowwidth=10,arrowinset=1,flip](400.379,-109.379)(79.622,-90.273,0.273)
 			\Arc[arrow,arrowpos=0.5,arrowlength=15,arrowwidth=10,arrowinset=1,flip,clock](399.621,-109.379)(79.622,-89.727,-180.273)
 			\end{picture}
 			&
 			\begin{picture}(181,195) (310,-191)
 			\SetWidth{0.0}
 			\SetColor{Black}
 			\Arc[arrow,arrowpos=0.75,arrowlength=15,arrowwidth=10,arrowinset=1,clock](400,-93)(80,90,-90)
 			\SetWidth{1.0}
 			\Arc[arrow,arrowpos=0.125,arrowlength=15,arrowwidth=10,arrowinset=1](400,-93)(79.981,132,492)
 			\Arc[arrow,arrowpos=0.5,arrowlength=15,arrowwidth=10,arrowinset=1](505.019,-93)(132.019,142.701,217.299)
 			\Arc[arrow,arrowpos=0.5,arrowlength=15,arrowwidth=10,arrowinset=1,clock](294.981,-93)(132.019,37.299,-37.299)
 			\CTri(383.447,-173)(400,-156.447)(416.553,-173){Black}{Black}\CTri(383.447,-173)(400,-189.553)(416.553,-173){Black}{Black}
 			\SetWidth{1.5}
 			\Line(385,2)(415,-28)
 			\Line(415,2)(385,-28)
 			\SetWidth{1.0}
 			\CBox(490.207,-85.207)(465.793,-60.793){Black}{Black}
 			\CBox(490.207,-125.207)(465.793,-100.793){Black}{Black}
 			\SetWidth{0.0}
 			\Arc[arrow,arrowpos=0.25,arrowlength=15,arrowwidth=10,arrowinset=1,clock](400,-93)(80,90,-90)
 			\end{picture}
 			&
 			\begin{picture}(183,195) (310,-191)
 			\SetWidth{1.0}
 			\SetColor{Black}
 			\CTri(383.447,-173)(400,-156.447)(416.553,-173){Black}{Black}\CTri(383.447,-173)(400,-189.553)(416.553,-173){Black}{Black}
 			\CBox(492.207,-105.207)(467.793,-80.793){Black}{Black}
 			\SetWidth{0.0}
 			\Arc[arrow,arrowpos=0.7,arrowlength=15,arrowwidth=10,arrowinset=1,clock](400,-93)(80,90,-90)
 			\SetWidth{1.0}
 			\Arc[arrow,arrowpos=0.125,arrowlength=15,arrowwidth=10,arrowinset=1](400,-93)(79.981,132,492)
 			\Arc[arrow,arrowpos=0.5,arrowlength=15,arrowwidth=10,arrowinset=1](505.019,-93)(132.019,142.701,217.299)
 			\Arc[clock](294.981,-93)(132.019,37.299,-37.299)
 			\SetWidth{1.5}
 			\Line(385,2)(415,-28)
 			\Line(415,2)(385,-28)
 			\SetWidth{0.0}
 			\Arc[arrow,arrowpos=0.3,arrowlength=15,arrowwidth=10,arrowinset=1,clock](400,-93)(80,90,-90)
 			\Arc[arrow,arrowpos=0.25,arrowlength=15,arrowwidth=10,arrowinset=1,clock](294.981,-93)(132.019,37.299,-37.299)
 			\Arc[arrow,arrowpos=0.75,arrowlength=15,arrowwidth=10,arrowinset=1,clock](294.981,-93)(132.019,37.299,-37.299)
 			\SetWidth{1.0}
 			\CBox(439.207,-105.207)(414.793,-80.793){Black}{Black}
 			\end{picture}
 			\\
 			{\Huge $T^{(5)}_{1}$}
 			&
 			{\Huge $T^{(5)}_{2}$}
 			&
 			{\Huge $T^{(5)}_{3}$}
 			&
 			{\Huge $T^{(5)}_{4}$}
 			&
 			{\Huge $T^{(5)}_{5}$}
 			\\
 			&
 			&
 			&
 			&
 			\\
 			\begin{picture}(197,195) (310,-191)
 			\SetWidth{1.0}
 			\SetColor{Black}
 			\CTri(383.447,-173)(400,-156.447)(416.553,-173){Black}{Black}\CTri(383.447,-173)(400,-189.553)(416.553,-173){Black}{Black}
 			\SetWidth{0.0}
 			\Arc[arrow,arrowpos=0.7,arrowlength=15,arrowwidth=10,arrowinset=1,clock](400,-93)(80,90,-90)
 			\SetWidth{1.0}
 			\Arc[arrow,arrowpos=0.125,arrowlength=15,arrowwidth=10,arrowinset=1](400,-93)(79.981,132,492)
 			\Arc[arrow,arrowpos=0.5,arrowlength=15,arrowwidth=10,arrowinset=1](505.019,-93)(132.019,142.701,217.299)
 			\Arc[clock](294.981,-93)(132.019,37.299,-37.299)
 			\SetWidth{1.5}
 			\Line(385,2)(415,-28)
 			\Line(415,2)(385,-28)
 			\SetWidth{0.0}
 			\Arc[arrow,arrowpos=0.3,arrowlength=15,arrowwidth=10,arrowinset=1,clock](400,-93)(80,90,-90)
 			\Arc[arrow,arrowpos=0.25,arrowlength=15,arrowwidth=10,arrowinset=1,clock](294.981,-93)(132.019,37.299,-37.299)
 			\Arc[arrow,arrowpos=0.75,arrowlength=15,arrowwidth=10,arrowinset=1,clock](294.981,-93)(132.019,37.299,-37.299)
 			\Text(440,-73)[]{\Huge{\Black{$A$}}}
 			\Text(490,-113)[]{\Huge{\Black{$B$}}}
 			\Text(490,-73)[]{\Huge{\Black{$A$}}}
 			\Text(440,-113)[]{\Huge{\Black{$B$}}}
 			\end{picture}
 			&	
 			\begin{picture}(183,195) (310,-191)
 			\SetWidth{0.0}
 			\SetColor{Black}
 			\Arc[arrow,arrowpos=0.75,arrowlength=15,arrowwidth=10,arrowinset=1,clock](400,-93)(80,90,-90)
 			\SetWidth{1.0}
 			\Arc[arrow,arrowpos=0.125,arrowlength=15,arrowwidth=10,arrowinset=1](400,-93)(79.981,132,492)
 			\Arc[arrow,arrowpos=0.5,arrowlength=15,arrowwidth=10,arrowinset=1](505.019,-93)(132.019,142.701,217.299)
 			\Arc[arrow,arrowpos=0.5,arrowlength=15,arrowwidth=10,arrowinset=1,clock](294.981,-93)(132.019,37.299,-37.299)
 			\CTri(383.447,-173)(400,-156.447)(416.553,-173){Black}{Black}\CTri(383.447,-173)(400,-189.553)(416.553,-173){Black}{Black}
 			\SetWidth{1.5}
 			\Line(385,2)(415,-28)
 			\Line(415,2)(385,-28)
 			\SetWidth{1.0}
 			\CBox(492.207,-105.207)(467.793,-80.793){Black}{Black}
 			\SetWidth{0.0}
 			\Arc[arrow,arrowpos=0.25,arrowlength=15,arrowwidth=10,arrowinset=1,clock](400,-93)(80,90,-90)
 			\end{picture}
 			&
 			&
 			&
 			\\
 			{\Huge $T^{(5)}_{6}$}
 			&
 			{\Huge $T^{(5)}_{7}$}
 			&
 			&
 			&
 		\end{tabular*}
 	}
 	\caption{Contributions to $T^{(5)}$}
 	\label{fig2}
 \end{table}
 
 \begin{table}[t]
 	\setlength{\extrarowheight}{0.5cm}
 	\setlength{\tabcolsep}{24pt}
 	\hspace*{1cm}
 	\centering
 	\resizebox{6.5cm}{!}{
 		\begin{tabular*}{20cm}{cc}
 			\begin{picture}(182,182) (309,-197)
 			\SetWidth{1.0}
 			\SetColor{Black}
 			\CTri(345.447,-176)(362,-159.447)(378.553,-176){Black}{Black}\CTri(345.447,-176)(362,-192.553)(378.553,-176){Black}{Black}
 			\Arc[arrow,arrowpos=0.5,arrowlength=15,arrowwidth=10,arrowinset=1](400,-106)(80,-180,0)
 			\SetWidth{1.5}
 			\Line(423,-21)(453,-51)
 			\Line(453,-21)(423,-51)
 			\SetWidth{1.0}
 			\Arc[arrow,arrowpos=0.5,arrowlength=15,arrowwidth=10,arrowinset=1,flip,clock](400,-106)(80,-180,-360)
 			\Arc[arrow,arrowpos=0.125,arrowlength=15,arrowwidth=10,arrowinset=1,flip](400,-106)(79.981,132,492)
 			\Arc[arrow,arrowpos=0.48,arrowlength=15,arrowwidth=10,arrowinset=1,flip,clock](261.636,-106)(122.364,34.894,-34.894)
 			\Arc[arrow,arrowpos=0.48,arrowlength=15,arrowwidth=10,arrowinset=1](538.364,-106)(122.364,145.106,214.894)
 			\Arc[arrow,arrowpos=0.48,arrowlength=15,arrowwidth=10,arrowinset=1,flip](602,-106)(250,163.74,196.26)
 			\Arc[arrow,arrowpos=0.48,arrowlength=15,arrowwidth=10,arrowinset=1,clock](198,-106)(250,16.26,-16.26)
 			\SetWidth{0.0}
 			\Arc[arrow,arrowpos=0.48,arrowlength=15,arrowwidth=10,arrowinset=1,clock](400.667,-106)(79.333,61.928,-61.928)
 			\end{picture}
 			&
 			\begin{picture}(162,196) (319,-190)
 			\SetWidth{1.0}
 			\SetColor{Black}
 			\CTri(383.447,-172)(400,-155.447)(416.553,-172){Black}{Black}\CTri(383.447,-172)(400,-188.553)(416.553,-172){Black}{Black}
 			\SetWidth{1.5}
 			\Line(385,4)(415,-26)
 			\Line(415,4)(385,-26)
 			\SetWidth{1.0}
 			\Arc[arrow,arrowpos=0,arrowlength=15,arrowwidth=10,arrowinset=1](400,-92)(79.981,132,492)
 			\Arc[arrow,arrowpos=0.5,arrowlength=15,arrowwidth=10,arrowinset=1,clock](321.441,-13.441)(78.572,1.051,-91.051)
 			\Arc[arrow,arrowpos=0.5,arrowlength=15,arrowwidth=10,arrowinset=1](478.559,-13.441)(78.572,178.949,271.051)
 			\Arc[arrow,arrowpos=0.5,arrowlength=15,arrowwidth=10,arrowinset=1](321.441,-170.559)(78.572,-1.051,91.051)
 			\Arc[arrow,arrowpos=0.5,arrowlength=15,arrowwidth=10,arrowinset=1,clock](478.559,-170.559)(78.572,-178.949,-271.051)
 			\SetWidth{0.0}
 			\Arc[arrow,arrowpos=0.5,arrowlength=15,arrowwidth=10,arrowinset=1,clock](400.379,-91.621)(79.622,90.273,-0.273)
 			\Arc[arrow,arrowpos=0.5,arrowlength=15,arrowwidth=10,arrowinset=1](400.379,-92.379)(79.622,-90.273,0.273)
 			\Arc[arrow,arrowpos=0.5,arrowlength=15,arrowwidth=10,arrowinset=1,clock](399.621,-92.379)(79.622,-89.727,-180.273)
 			\end{picture}
 			\\
 			{\Huge $K^{(5)}_{1}$}
 			&
 			{\Huge $K^{(5)}_{2}$}
 		\end{tabular*}
 	}
 	\caption{Contributions to $K^{(5)}$}
 	\label{fig3}
 \end{table}
 \clearpage
 
 \begin{align}
 (\gamma^{(4)}_1)_i{}^j&=\Ybar_{ilmn}(Y_2)_k{}^lY^{kmnj},\nn
 (\gamma^{(4)}_2)_i{}^j&=\Ybar_{ipqr}Y^{pqmn}\Ybar_{mnkl}Y^{klrj},\nn
 \gamma^{(4)}_3&=Y_2C_RC_R,\nn
 (\gamma^{(4)}_4)_i{}^j&=\Ybar_{ikln}(C_RC_R)^n{}_mY^{klmj}\nn
(\gamma^{(4)}_{5})_i{}^j&=\Ybar_{iklm}(C_R)_n{}^mY^{pkln}(C_R)_p{}^j,\nn
 (\gamma^{(4)}_{6})_i{}^j&=\Ybar_{iklm}(C_R)_n{}^l(C_R)_p{}^mY^{knpj},\nn
 (\gamma^{(4)}_7)_i{}^j&=\Ybar_{iklm}(R_AR_B)_n{}^mY^{pkln}(R_BR_A)_p{}^j,\nn
 (\gamma^{(4)}_8)_i{}^j&=\Ybar_{iklm}(R_AR_B)_n{}^l(R_AR_B)_p{}^mY^{knpj},\nn
 \gamma^{(4)}_9&=Y_2C_R,\nn
 (\gamma^{(4)}_{10})_i{}^j&=\Ybar_{iklm}(C_R)_n{}^mY^{klnj},\nn
 (\gamma^{(4)}_{11})_i{}^j&=\tfrac12\tr(Y_2R_AR_B)(\{R_A,R_B\})_i{}^j,\nn
  (\gamma^{(4)}_{12})_i{}^j&=\Ybar_{iklm}(R_AC_R)_n{}^l(R_A)_p{}^mY^{knpj}.
 \label{Ystruct}
 \end{align}
The structures $\gamma^{(4)}_{1-11}$ form a basis for 2nd rank tensors with four gauge matrices; $\gamma^{(4)}_{12}$ is not independent but has been retained since it appears naturally in diagrammatic calculations (and in fact ultimately cancels). The ellipsis in Eq.~\eqref{gdef} indicates Yukawa-independent terms which we have not computed.
We then find that Eq.~\eqref{grad} entails
\begin{align}
a_{1}^{(7)}+\tfrac13\ta=&t^{(5)}_1=2g^{(4)}_1+\tfrac12(t^{(5)}_2+k^{(5)}_1),\nn
a_{2}^{(7)}=&4g^{(4)}_2=\tfrac23t^{(5)}_3=\tfrac23t^{(5)}_3+\tfrac43k^{(5)}_2,\nn
a_{3}^{(7)}+\tfrac19\ta=&\tfrac19(t^{(5)}_1+t^{(5)}_2+k^{(5)}_1),\nn
a_{4}^{(7)}-\tfrac43\ta=&2g^{(4)}_3-t^{(5)}_1-t^{(5)}_2-k^{(5)}_1+\tfrac16t^{(5)}_4,\nn
a_{5}^{(7)}-4\ta=&4g^{(4)}_4-6t^{(5)}_2-6k^{(5)}_1=-6t^{(5)}_1+t^{(5)}_4,\nn
a_{6}^{(7)}=&4g^{(4)}_{5}=\tfrac23t^{(5)}_5,\nn
a_{7}^{(7)}=&4g^{(4)}_{6}=\tfrac23t^{(5)}_5,\nn
a_{8}^{(7)}=&4g^{(4)}_7=\tfrac23t^{(5)}_6,\nn
a_{9}^{(7)}=&4g^{(4)}_8=\tfrac23t^{(5)}_6,\nn
a_{10}^{(7)}=&-4t^{(5)}_3=-4t^{(5)}_3-8k^{(5)}_2,\nn
a_{11}^{(7)}=&-2\Ttil t^{(5)}_3=-2\Ttil t^{(5)}_3-4\Ttil k^{(5)}_2,\nn
a_{12}^{(7)}-\tfrac23\Ttil\ta=&2g^{(4)}_9-\tfrac12\Ttil(t^{(5)}_1+t^{(5)}_2+k^{(5)}_1)+\tfrac16t^{(5)}_7,\nn
a_{13}^{(7)}-2\Ttil\ta=&4g^{(4)}_{10}-3\Ttil (t^{(5)}_2+k^{(5)}_1)=-3\Ttil t^{(5)}_1+t^{(5)}_7,\nn
a_{14}^{(7)}=&2g^{(4)}_{11},
\label{aeqs}
\end{align}
where $\ta$ is the parameter introduced in Eq.~\eqref{adef}.
The $a$-function coefficients $a_{2}^{(7)}$, $a_{6-9}^{(7)}$,  $a_{14}^{(7)}$, are given directly in terms of the anomalous dimension coefficients in Eq.~\eqref{aeqs}; while the remaining ones are given by
\begin{align}
a_{1}^{(7)}=&\tfrac43g^{(4)}_1,\nn
a_{3}^{(7)}=&0,\nn
a_{4}^{(7)}=&\tfrac83g^{(4)}_1+2g^{(4)}_3+\tfrac23g^{(4)}_4,\nn
a_{5}^{(7)}=&8g^{(4)}_1+4g^{(4)}_4,\nn
a_{10}^{(7)}=&-24g^{(4)}_2,\nn
a_{11}^{(7)}=&-12\Ttil g^{(4)}_2,\nn
a_{12}^{(7)}=&\tfrac43\Ttil g^{(4)}_1+2g^{(4)}_9+\tfrac23g^{(4)}_{10},\nn
a_{13}^{(7)}=&4\Ttil g^{(4)}_1+4g^{(4)}_{10},
\label{acoeffs}
\end{align}
together with
\begin{align}
t_1^{(5)}=&\tfrac13(4g^{(4)}_1+\ta ),\nn
t_2^{(5)}+k^{(5)}_1=&\tfrac23(-2g^{(4)}_1+\ta ),\nn
t_3^{(6)}=6g^{(4)}_2,\nn
t_4^{(5)}=&16g^{(4)}_1+4g^{(4)}_4-2\ta ,\nn
t_5^{(5)}=&6g^{(4)}_{5},\nn
t_6^{(5)}=&6g^{(4)}_7,\nn
t_7^{(5)}=&8\Ttil g^{(4)}_1+4g^{(4)}_{10}-\Ttil \ta,\nn
k_2^{(5)}=&0,
\end{align}
subject to the consistency conditions
\be
 g^{(4)}_5=g^{(4)}_{6},\quad g^{(4)}_{7}=g^{(4)}_{8}.
\label{fourcon}
\ee
Turning to the  Yukawa-independent terms, it is clear that we may satisfy Eq.~\eqref{grad} if for each Yukawa-independent term $X^i{}_j$, we add to $A^{(7)}$ a term $Y^{ijkl}X^m{}_{(i}Y_{jkl)m}$, and therefore there will be no further constraints.

As we have observed already in four dimensions\cite{OsbJacnew,JacPoole} and six dimensions\cite{asix}, and indeed at lower orders in three dimensions\cite{JJP,JPa}, the $a$-function coefficients are determined completely (within a given renormalisation scheme) up to the arbitrariness parametrised by the coefficient $\ta$.

\section{Four-loop calculation}
In this section we describe the diagrammatic computation of the four-loop anomalous dimension, again focussing on contributions containing Yukawa couplings. We are therefore concerned with the calculation of four-loop two-point diagrams.
Two large classes
 of diagrams may be immediately discarded as giving no contribution to the anomalous dimension
 \cite{GN}. The first
 consists of those diagrams in which the first (last) vertex encountered along
 the incoming (outgoing) chiral line has a single gauge line. In this case after performing the superspace $D$-algebra\footnote{See Appendix A for conventions and definitions.} one is left with a diagram which is finite by power counting.  These diagrams are shown
 schematically in Table~\ref{fig4}(a).
 The second class consists of those diagrams which
 contain a one-loop subdiagram with one gauge and one chiral line, depicted
 in Table~\ref{fig4}(b); in this case one finds that one is left with fewer than two $D$'s and two $\Dbar$'s on the loop shown, hence giving a vanishing contribution. 
 
 \begin{table}[t]
 	\setlength{\extrarowheight}{0.5cm}
 	\setlength{\tabcolsep}{24pt}
 	\hspace*{0cm}
 	\centering
 	\resizebox{9cm}{!}{
 		\begin{tabular*}{20cm}{cc}
 			\begin{picture}(226,105) (287,-229)
 			\SetWidth{1.0}
 			\SetColor{Black}
 			\Line[arrow,arrowpos=0.5,arrowlength=15,arrowwidth=10,arrowinset=1](288,-183)(320,-183)
 			\Line[arrow,arrowpos=0.4,arrowlength=15,arrowwidth=10,arrowinset=1](416,-183)(512,-183)
 			\Arc[clock](368,-186.1)(48.1,176.305,3.695)
 			\Arc(368,-179.9)(48.1,-176.305,-3.695)
 			\Line(344,-147)(388,-224)
 			\Line(350,-143)(394,-220)
 			\Line(356,-140)(399,-215)
 			\Line(364,-140)(404,-210)
 			\Line(372,-139)(409,-204)
 			\Line(380,-140)(412,-196)
 			\Line(391,-144)(415,-187)
 			\Line(337,-149)(381,-226)
 			\Line(332,-154)(374,-228)
 			\Line(328,-161)(366,-228)
 			\Line(324,-168)(358,-227)
 			\Line(320,-177)(347,-223)
 			\Line(323,-195)(331,-211)
 			\Line(399,-150)(356,-224)
 			\Line(392,-147)(348,-224)
 			\Line(387,-143)(342,-221)
 			\Line(337,-216)(380,-140)
 			\Line(372,-140)(332,-211)
 			\Line(328,-203)(364,-140)
 			\Line(357,-140)(323,-199)
 			\Line(321,-189)(347,-142)
 			\Line(404,-154)(362,-227)
 			\Line(408,-161)(369,-228)
 			\Line(412,-168)(379,-227)
 			\Line(415,-178)(388,-224)
 			\Line(412,-196)(400,-216)
 			\PhotonArc[clock](406.618,-214.667)(84.537,114.919,21.999){5}{9.5}
 			\end{picture}
 			&
 			\begin{picture}(226,141) (287,-193)
 			\SetWidth{1.0}
 			\SetColor{Black}
 			\Line[arrow,arrowpos=0.5,arrowlength=15,arrowwidth=10,arrowinset=1](288,-147)(350,-147)
 			\Line[arrow,arrowpos=0.5,arrowlength=15,arrowwidth=10,arrowinset=1](448,-147)(512,-147)
 			\Arc[clock](399,-150.1)(48.1,176.305,3.695)
 			\Arc(399,-143.9)(48.1,-176.305,-3.695)
 			\Line(375,-111)(419,-188)
 			\Line(381,-107)(425,-184)
 			\Line(387,-104)(430,-179)
 			\Line(395,-104)(435,-174)
 			\Line(403,-103)(440,-168)
 			\Line(411,-104)(443,-160)
 			\Line(422,-108)(446,-151)
 			\Line(368,-113)(412,-190)
 			\Line(363,-118)(405,-192)
 			\Line(359,-125)(397,-192)
 			\Line(355,-132)(389,-191)
 			\Line(351,-141)(378,-187)
 			\Line(354,-159)(362,-175)
 			\Line(430,-114)(387,-188)
 			\Line(423,-111)(379,-188)
 			\Line(418,-107)(373,-185)
 			\Line(368,-180)(411,-104)
 			\Line(403,-104)(363,-175)
 			\Line(359,-167)(395,-104)
 			\Line(388,-104)(354,-163)
 			\Line(352,-153)(378,-106)
 			\Line(435,-118)(393,-191)
 			\Line(439,-125)(400,-192)
 			\Line(443,-132)(410,-191)
 			\Line(446,-142)(419,-188)
 			\Line(443,-160)(431,-180)
 			\Line[arrow,arrowpos=0.5,arrowlength=15,arrowwidth=10,arrowinset=1](367,-113)(398,-64)
 			\Line[arrow,arrowpos=0.5,arrowlength=15,arrowwidth=10,arrowinset=1](398,-64)(433,-116)
 			\PhotonArc(380.81,-87.431)(29.06,53.735,241.626){5}{7.5}
 			\end{picture}
 			\\
 			{\Huge $\left( a \right)$}
 			&
 			{\Huge $\left( b \right)$}
 		\end{tabular*}
 	}
 	\caption{Classes of diagrams that do not contribute}
 	\label{fig4}
 \end{table}




The diagrams which do potentially give non-trivial contributions to the four-loop anomalous dimension are depicted
in Table~\ref{diagfoura}. With the exception of \ref{diagfoura}(h) (which will be discussed in more detail shortly),  
 the momentum integrals obtained after performing the superspace $D$-algebra in these diagrams may be expressed using integration by parts in terms of a relatively small basis of momentum integrals\cite{MSS,MSSa}
 which are depicted in Table~\ref{diagfourc},
 and whose divergences are also listed in Appendix B. (The graph labelled $X$ in Table~\ref{diagfourc} will also be discussed in more detail shortly.) 
The massless 4-loop 2-point functions depicted in Table 6 are assumed to have their UV subdivergences subtracted; the  ``dot'' on the propagator in $\tilde Y$ in Table~\ref{diagfourc} represents a double propagator. The results
 given later, and also most of these conventions for labelling the diagrams,
 are taken from Refs.~\cite{MSS,MSSa}; except for $\tilde Y$ which was defined (without the tilde, used here to avoid confusion with the Yukawa coupling)  in Ref.~\cite{JPa}.
Our results for the diagrams of Table~\ref{diagfoura} are listed in Table~\ref{restaba}.  The central columns of Table~\ref{restaba} show the divergent contribution from each diagram (again, except \ref{diagfoura}(h)) expressed in terms of this basis. These momentum integrals multiply a variety of group structures, as defined in Eqs.~\eqref{defs}, \eqref{Ystruct},
 which are tabulated in the final column of Table~\ref{restaba}.
Finally the first column of Table~\ref{restaba} contains
 an overall symmetry factor. The resulting contribution to the
 two-point function for each diagram is therefore obtained
 by adding the momentum integrals with the coefficients listed in the appropriate
 row and multiplying the resulting sum by the corresponding symmetry factor and
 group structure. For instance, row (j) of Table~\ref{restaba} denotes
 a contribution
 \be
 (+1)(-2I_4+I_{4bbb})\left(\gamma^{(4)}_7-\tfrac{1}{12}C_G\gamma^{(4)}_9\right).
 \ee
We note here the cancellation of $\gamma_{12}^{(4)}$ between rows (e) and (i), as mentioned in the previous Section. 

We now return to graph (h) of Table $5$. After performing the superspace
$D$-algebra, this results in the momentum integral labelled $X$ in Table $6$.
In it there is an implicit spinor trace over the momenta of the rim
propagators, $\kslash$, where we use three dimensional $\gamma$-matrices with
$\mbox{Tr} I$~$=$~$2$. This particular integral provided us with a technical
challenge compared to the other graphs we had to evaluate. Accordingly we describe 
our method for evaluating it in more detail than usual. Firstly, by power
counting it is straightforward to see that the graph is primitively divergent,
which provides a shortcut to finding the divergence. Either we can reroute the
external momentum through the diagram in such a way that it becomes simpler to
compute, or we can use a vacuum bubble expansion, such as that discussed in
Refs.~\cite{1,2}. In the former case one has to be careful that infrared divergences
are not introduced. However, we have chosen to follow the latter course as it
is more systematic and accessible to recently developed integration-by-parts
algorithms. In converting the massless four-loop $2$-point function to vacuum
bubbles we recursively apply the identity
\begin{equation}
\frac{1}{(k-p)^2} ~=~ \frac{1}{[k^2+m^2]} ~+~
\frac{2kp - p^2 + m^2}{(k-p)^2[k^2+m^2]}
\end{equation}
to all propagators, where $k$ can be regarded as a loop momentum. The recursion
terminates when all resulting integrals involving the external momentum are
finite by Weinberg's theorem. In our case one in effect replaces the scalar
propagators of the graph by the first term of the identity. What remains is an
integral with products of scalar products of internal and external momenta
after the trace has been taken. To proceed we have applied the Laporta
algorithm\cite{3}. This constructs identities based on integration by parts
for all such scalar product integrals and then reduces these to a base set of
what is termed master integrals. Specifically these are four-loop massive
vacuum diagrams. In particular they have been evaluated numerically to very
high precision in three dimensions in Ref.~\cite{4} using the approach of Ref.~\cite{5},
running parallel to the same calculation in four dimensions\cite{6}. Therefore
for our particular graph we have constructed a database of relations between
all possible integrals within that of graph $X$ of Table $6$ using the
{\sc Reduze} formulation\cite{7} of the Laporta algorithm. We have used
{\sc Form}\cite{8,9} to handle the resulting algebra. One aspect of the
integration-by-parts approach is that one has to substitute terms from the
masters beyond the poles in $\epsilon$. This is because factors of $1/(d-3)$
will appear as coefficients of the master integrals in the decomposition of the
original integral. In addition, as several of the masters have double poles in
$\epsilon$ then in order to have an answer which corresponds to a primitive the
poles higher than the simple one have to cancel. It is reassuring to find that
this is indeed the case when we perform the actual computation, giving us a
strong check. Interestingly it transpires that integral $X$ of Table $6$ is in
fact finite.

\begin{table}[t]
	\setlength{\extrarowheight}{1cm}
	\setlength{\tabcolsep}{24pt}
	\hspace*{-6.5cm}
	\centering
	\resizebox{7cm}{!}{
		\begin{tabular*}{20cm}{cccc}
			\begin{picture}(226,182) (287,-197)
			\SetWidth{1.0}
			\SetColor{Black}
			\Arc[arrow,arrowpos=0.5,arrowlength=15,arrowwidth=10,arrowinset=1,flip](400,-102.9)(48.1,-176.305,-3.695)
			\Arc[arrow,arrowpos=0.5,arrowlength=15,arrowwidth=10,arrowinset=1,flip,clock](400,-109.1)(48.1,176.305,3.695)
			\Arc[arrow,arrowpos=0.5,arrowlength=15,arrowwidth=10,arrowinset=1,clock](400,-106)(80,-180,-360)
			\Arc[arrow,arrowpos=0.5,arrowlength=15,arrowwidth=10,arrowinset=1](400,-106)(80,-180,0)
			\Line[arrow,arrowpos=0.5,arrowlength=15,arrowwidth=10,arrowinset=1,flip](352,-106)(448,-106)
			\Line[arrow,arrowpos=0.5,arrowlength=15,arrowwidth=10,arrowinset=1](320,-106)(352,-106)
			\Line[arrow,arrowpos=0.5,arrowlength=15,arrowwidth=10,arrowinset=1](448,-106)(480,-106)
			\Line(288,-106)(320,-106)
			\Line(480,-106)(512,-106)
			\end{picture}
			&
			\begin{picture}(226,182) (287,-197)
			\SetWidth{1.0}
			\SetColor{Black}
			\Arc[arrow,arrowpos=0.5,arrowlength=15,arrowwidth=10,arrowinset=1,flip,clock](400,-119.2)(93.2,116.785,63.215)
			\Arc[arrow,arrowpos=0.5,arrowlength=15,arrowwidth=10,arrowinset=1](400,-106)(80,-180,0)
			\Line(288,-106)(320,-106)
			\Line(480,-106)(512,-106)
			\Arc[arrow,arrowpos=0.5,arrowlength=15,arrowwidth=10,arrowinset=1,clock](408.391,-108.67)(88.432,178.27,124.739)
			\Arc[arrow,arrowpos=0.5,arrowlength=15,arrowwidth=10,arrowinset=1](320.352,-60.877)(45.125,-90.446,33.455)
			\Arc[arrow,arrowpos=0.5,arrowlength=15,arrowwidth=10,arrowinset=1,clock](391.609,-108.67)(88.432,55.261,1.73)
			\Arc[arrow,arrowpos=0.5,arrowlength=15,arrowwidth=10,arrowinset=1](479.648,-60.877)(45.125,146.545,270.446)
			\Arc[arrow,arrowpos=0.5,arrowlength=15,arrowwidth=10,arrowinset=1,flip](400,-13.22)(47.78,-151.525,-28.475)
			\end{picture}
			&
			\begin{picture}(226,182) (287,-197)
			\SetWidth{1.0}
			\SetColor{Black}
			\Arc[arrow,arrowpos=0.5,arrowlength=15,arrowwidth=10,arrowinset=1,clock](400,-106)(80,-180,-360)
			\Arc[arrow,arrowpos=0.5,arrowlength=15,arrowwidth=10,arrowinset=1](400,-106)(80,-180,0)
			\Line(288,-106)(320,-106)
			\Line(480,-106)(512,-106)
			\Line[arrow,arrowpos=0.5,arrowlength=15,arrowwidth=10,arrowinset=1](320,-106)(480,-106)
			\PhotonArc(400,-106)(50.249,96,456){5}{20}
			\end{picture}
			&
			\begin{picture}(226,194) (287,-185)
			\SetWidth{1.0}
			\SetColor{Black}
			\Arc[arrow,arrowpos=0.75,arrowlength=15,arrowwidth=10,arrowinset=1,clock](400,-94)(80,-180,-360)
			\Arc[arrow,arrowpos=0.5,arrowlength=15,arrowwidth=10,arrowinset=1](400,-94)(80,-180,0)
			\Line(288,-94)(320,-94)
			\Line(480,-94)(512,-94)
			\Line[arrow,arrowpos=0.5,arrowlength=15,arrowwidth=10,arrowinset=1](320,-94)(480,-94)
			\PhotonArc[clock](395.512,-42.709)(45.931,161.323,25.41){5}{5.5}
			\PhotonArc[clock](335.028,-17.066)(48.972,0.077,-101.816){5}{5.5}
			\end{picture}
			\\
			{\Huge $\left( a \right)$}
			&
			{\Huge $\left( b \right)$}
			&
			{\Huge $\left( c \right)$}
			&
			{\Huge $\left( d \right)$}
			\\
			&
			&
			&
			\\
			\begin{picture}(226,195) (287,-184)
			\SetWidth{1.0}
			\SetColor{Black}
			\Arc[arrow,arrowpos=0.25,arrowlength=15,arrowwidth=10,arrowinset=1,clock](400,-93)(80,-180,-360)
			\Arc[arrow,arrowpos=0.5,arrowlength=15,arrowwidth=10,arrowinset=1](400,-93)(80,-180,0)
			\Line(288,-93)(320,-93)
			\Line(480,-93)(512,-93)
			\Line[arrow,arrowpos=0.25,arrowlength=15,arrowwidth=10,arrowinset=1](320,-93)(480,-93)
			\PhotonArc[clock](400,-37.086)(42.086,161.884,18.116){5}{6.5}
			\Photon(400,-13)(400,-93){5}{6}
			\end{picture}
			&
			\begin{picture}(226,172) (287,-207)
			\SetWidth{1.0}
			\SetColor{Black}
			\Arc[arrow,arrowpos=0.25,arrowlength=15,arrowwidth=10,arrowinset=1,clock](400,-116)(80,-180,-360)
			\Arc[arrow,arrowpos=0.5,arrowlength=15,arrowwidth=10,arrowinset=1](400,-116)(80,-180,0)
			\Line(288,-116)(320,-116)
			\Line(480,-116)(512,-116)
			\Line[arrow,arrowpos=0.5,arrowlength=15,arrowwidth=10,arrowinset=1](320,-116)(480,-116)
			\Photon(400,-36)(360,-116){5}{7.5}
			\Photon(400,-36)(440,-116){5}{7.5}
			\end{picture}
			&
			\begin{picture}(226,174) (287,-205)
			\SetWidth{1.0}
			\SetColor{Black}
			\Arc[arrow,arrowpos=0.25,arrowlength=15,arrowwidth=10,arrowinset=1,clock](400,-114)(80,-180,-360)
			\Arc[arrow,arrowpos=0.5,arrowlength=15,arrowwidth=10,arrowinset=1](400,-114)(80,-180,0)
			\Line(288,-114)(320,-114)
			\Line(480,-114)(512,-114)
			\Line[arrow,arrowpos=0.25,arrowlength=15,arrowwidth=10,arrowinset=1](320,-114)(480,-114)
			\PhotonArc(430,-74)(50,126.87,233.13){5}{6.5}
			\PhotonArc[clock](370,-74)(50,53.13,-53.13){5}{6.5}
			\end{picture}
			&
			\begin{picture}(226,172) (287,-207)
			\SetWidth{1.0}
			\SetColor{Black}
			\Arc[arrow,arrowpos=0.25,arrowlength=15,arrowwidth=10,arrowinset=1,clock](400,-116)(80,-180,-360)
			\Arc[arrow,arrowpos=0.5,arrowlength=15,arrowwidth=10,arrowinset=1](400,-116)(80,-180,0)
			\Line(288,-116)(320,-116)
			\Line(480,-116)(512,-116)
			\Line[arrow,arrowpos=0.5,arrowlength=15,arrowwidth=10,arrowinset=1](320,-116)(480,-116)
			\Photon(362,-46)(440,-116){5}{6}
			\Photon(440,-46)(362,-116){5}{6}
			\end{picture}
			\\
			{\Huge $\left( e \right)$}
			&
			{\Huge $\left( f \right)$}
			&
			{\Huge $\left( g \right)$}
			&
			{\Huge $\left( h \right)$}
			\\
			&
			&
			&
			\\
			\begin{picture}(226,162) (287,-207)
			\SetWidth{1.0}
			\SetColor{Black}
			\Arc[arrow,arrowpos=0.25,arrowlength=15,arrowwidth=10,arrowinset=1,clock](400,-126)(80,-180,-360)
			\Arc[arrow,arrowpos=0.25,arrowlength=15,arrowwidth=10,arrowinset=1](400,-126)(80,-180,0)
			\Line(288,-126)(320,-126)
			\Line(480,-126)(512,-126)
			\Line[arrow,arrowpos=0.25,arrowlength=15,arrowwidth=10,arrowinset=1](320,-126)(480,-126)
			\Photon(400,-46)(400,-206){5}{12}
			\end{picture}
			&
			\begin{picture}(226,116) (287,-223)
			\SetWidth{1.0}
			\SetColor{Black}
			\Arc[arrow,arrowpos=0.25,arrowlength=15,arrowwidth=10,arrowinset=1,clock](370,-172)(50,-180,-360)
			\Arc[arrow,arrowpos=0.25,arrowlength=15,arrowwidth=10,arrowinset=1](370,-172)(50,-180,0)
			\Line(288,-172)(320,-172)
			\Line(480,-172)(512,-172)
			\Line[arrow,arrowpos=0.25,arrowlength=15,arrowwidth=10,arrowinset=1](320,-172)(480,-172)
			\PhotonArc[clock](411,-202.44)(90.286,117.008,19.703){5}{6.5}
			\PhotonArc[clock](426.674,-256.78)(109.516,96.646,50.727){5}{4.5}
			\end{picture}
			&
			\begin{picture}(226,127) (287,-212)
			\SetWidth{1.0}
			\SetColor{Black}
			\Arc[arrow,arrowpos=0.25,arrowlength=15,arrowwidth=10,arrowinset=1,clock](370,-161)(50,-180,-360)
			\Arc[arrow,arrowpos=0.25,arrowlength=15,arrowwidth=10,arrowinset=1](370,-161)(50,-180,0)
			\Line(288,-161)(320,-161)
			\Line(480,-161)(512,-161)
			\Line[arrow,arrowpos=0.25,arrowlength=15,arrowwidth=10,arrowinset=1](320,-161)(480,-161)
			\PhotonArc[clock](421.407,-165.214)(74.712,133.478,3.233){5}{7.5}
			\PhotonArc[clock](403.591,-210.11)(104.648,108.723,27.988){5}{6.5}
			\end{picture}
			&
			\begin{picture}(226,140) (287,-218)
			\SetWidth{1.0}
			\SetColor{Black}
			\Arc[arrow,arrowpos=0.25,arrowlength=15,arrowwidth=10,arrowinset=1,clock](370,-148)(50,-180,-360)
			\Arc[arrow,arrowpos=0.25,arrowlength=15,arrowwidth=10,arrowinset=1](370,-148)(50,-180,0)
			\Line(288,-148)(320,-148)
			\Line(480,-148)(512,-148)
			\Line[arrow,arrowpos=0.25,arrowlength=15,arrowwidth=10,arrowinset=1](320,-148)(480,-148)
			\PhotonArc[clock](411.638,-152.397)(68.503,127.432,3.68){5}{6.5}
			\PhotonArc(411.638,-143.603)(68.503,-127.432,-3.68){5}{6.5}
			\end{picture}
			\\
			{\Huge $\left( i \right)$}
			&
			{\Huge $\left( j \right)$}
			&
			{\Huge $\left( k \right)$}
			&
			{\Huge $\left( l \right)$}
			\\
			&
			&
			&
			\\
			\begin{picture}(226,172) (287,-202)
			\SetWidth{1.0}
			\SetColor{Black}
			\Arc[arrow,arrowpos=0.5,arrowlength=15,arrowwidth=10,arrowinset=1,clock](400,-116)(40,-180,-360)
			\Line(288,-116)(320,-116)
			\Line(480,-116)(512,-116)
			\Line[arrow,arrowpos=0.5,arrowlength=15,arrowwidth=10,arrowinset=1](320,-116)(480,-116)
			\Arc[arrow,arrowpos=0.5,arrowlength=15,arrowwidth=10,arrowinset=1](400,-116)(40,-180,0)
			\PhotonArc[clock](400,-116)(80,-180,-360){5}{20.5}
			\PhotonArc(400,-116)(80,-180,0){5}{20.5}
			\end{picture}
			&
			\begin{picture}(226,172) (287,-207)
			\SetWidth{1.0}
			\SetColor{Black}
			\Arc[arrow,arrowpos=0.25,arrowlength=15,arrowwidth=10,arrowinset=1,clock](400,-116)(80,-180,-360)
			\Arc[arrow,arrowpos=0.5,arrowlength=15,arrowwidth=10,arrowinset=1](400,-116)(80,-180,0)
			\Line(288,-116)(320,-116)
			\Line(480,-116)(512,-116)
			\Line[arrow,arrowpos=0.25,arrowlength=15,arrowwidth=10,arrowinset=1](320,-116)(480,-116)
			\Photon(400,-36)(400,-116){5}{6.5}
			\Arc(400,-75)(16.401,142,502)
			\Line(389,-63)(412,-84)
			\Line(412,-64)(389,-85)
			\Line(407,-61)(385,-81)
			\Line(414,-69)(392,-89)
			\Line(401,-59)(384,-75)
			\Line(416,-73)(396,-91)
			\Line(416,-78)(400,-92)
			\Line(385,-67)(409,-89)
			\Line(394,-59)(416,-79)
			\Line(402,-60)(416,-73)
			\Line(384,-74)(403,-90)
			\Line(386,-81)(398,-92)
			\end{picture}
			&
			\begin{picture}(226,182) (287,-197)
			\SetWidth{1.0}
			\SetColor{Black}
			\Arc[arrow,arrowpos=0.5,arrowlength=15,arrowwidth=10,arrowinset=1,clock](400,-106)(80,-180,-360)
			\Arc[arrow,arrowpos=0.5,arrowlength=15,arrowwidth=10,arrowinset=1](400,-106)(80,-180,0)
			\Line(288,-106)(320,-106)
			\Line(480,-106)(512,-106)
			\Line[arrow,arrowpos=0.5,arrowlength=15,arrowwidth=10,arrowinset=1](320,-106)(480,-106)
			\Arc(400,-65)(16.401,142,502)
			\Line(389,-53)(412,-74)
			\Line(412,-54)(389,-75)
			\Line(407,-51)(385,-71)
			\Line(414,-59)(392,-79)
			\Line(401,-49)(384,-65)
			\Line(416,-63)(396,-81)
			\Line(416,-68)(400,-82)
			\Line(385,-57)(409,-79)
			\Line(394,-49)(416,-69)
			\Line(402,-50)(416,-63)
			\Line(384,-64)(403,-80)
			\Line(386,-71)(398,-82)
			\PhotonArc(399,-4.435)(61.573,-141.22,-38.78){5}{10.5}
			\end{picture}
			&
			\begin{picture}(226,172) (287,-207)
			\SetWidth{1.0}
			\SetColor{Black}
			\Arc[arrow,arrowpos=0.25,arrowlength=15,arrowwidth=10,arrowinset=1,clock](400,-116)(80,-180,-360)
			\Arc[arrow,arrowpos=0.5,arrowlength=15,arrowwidth=10,arrowinset=1](400,-116)(80,-180,0)
			\Line(288,-116)(320,-116)
			\Line(480,-116)(512,-116)
			\Line[arrow,arrowpos=0.5,arrowlength=15,arrowwidth=10,arrowinset=1](320,-116)(480,-116)
			\Arc(400,-88)(16.401,142,502)
			\Line(389,-76)(412,-97)
			\Line(412,-77)(389,-98)
			\Line(407,-74)(385,-94)
			\Line(414,-82)(392,-102)
			\Line(401,-72)(384,-88)
			\Line(416,-86)(396,-104)
			\Line(416,-91)(400,-105)
			\Line(385,-80)(409,-102)
			\Line(394,-72)(416,-92)
			\Line(402,-73)(416,-86)
			\Line(384,-87)(403,-103)
			\Line(386,-94)(398,-105)
			\PhotonArc(400,-66)(26.401,-37,323){3}{12}
			\end{picture}
			\\
			{\Huge $\left( m \right)$}
			&
			{\Huge $\left( n \right)$}
			&
			{\Huge $\left( o \right)$}
			&
			{\Huge $\left( p \right)$}
			\\
			&
			&
			&
			\\
			\begin{picture}(226,162) (287,-137)
			\SetWidth{1.0}
			\SetColor{Black}
			\Line[arrow,arrowpos=0.5,arrowlength=15,arrowwidth=10,arrowinset=1](288,-126)(400,-126)
			\Line[arrow,arrowpos=0.5,arrowlength=15,arrowwidth=10,arrowinset=1](400,-126)(512,-126)
			\Arc[arrow,arrowpos=0.5,arrowlength=15,arrowwidth=10,arrowinset=1,flip](400,-36)(60,90,270)
			\Arc[arrow,arrowpos=0.5,arrowlength=15,arrowwidth=10,arrowinset=1,flip,clock](400,-36)(60,90,-90)
			\Arc[arrow,arrowpos=0.5,arrowlength=15,arrowwidth=10,arrowinset=1,flip,clock](320,-36)(100,36.87,-36.87)
			\Arc[arrow,arrowpos=0.5,arrowlength=15,arrowwidth=10,arrowinset=1,flip](480,-36)(100,143.13,216.87)
			\PhotonArc(400,-63.924)(56.076,177.011,362.989){5}{12.5}
			\end{picture}
			&
			\begin{picture}(226,172) (287,-127)
			\SetWidth{1.0}
			\SetColor{Black}
			\Line[arrow,arrowpos=0.5,arrowlength=15,arrowwidth=10,arrowinset=1](288,-116)(400,-116)
			\Line[arrow,arrowpos=0.5,arrowlength=15,arrowwidth=10,arrowinset=1](400,-116)(512,-116)
			\PhotonArc(400,-53.924)(56.076,177.011,362.989){5}{12.5}
			\Arc[arrow,arrowpos=0.5,arrowlength=15,arrowwidth=10,arrowinset=1,clock](400,-26)(60,-180,-360)
			\Arc[arrow,arrowpos=0.5,arrowlength=15,arrowwidth=10,arrowinset=1](400,-26)(60,-180,0)
			\Arc[arrow,arrowpos=0.5,arrowlength=15,arrowwidth=10,arrowinset=1,clock](400,-106)(100,126.87,53.13)
			\Arc[arrow,arrowpos=0.5,arrowlength=15,arrowwidth=10,arrowinset=1](400,54)(100,-126.87,-53.13)
			\end{picture}
			&
			&
			\\
			{\Huge $\left( q \right)$}
			&
			{\Huge $\left( r \right)$}
			&
			&
			\\
			&
			&
			&
			\\
			\begin{picture}(226,182) (287,-197)
			\SetWidth{1.0}
			\SetColor{Black}
			\Arc[arrow,arrowpos=0.5,arrowlength=15,arrowwidth=10,arrowinset=1,clock](400,-106)(80,-180,-360)
			\Arc[arrow,arrowpos=0.5,arrowlength=15,arrowwidth=10,arrowinset=1](400,-106)(80,-180,0)
			\Line(288,-106)(320,-106)
			\Line(480,-106)(512,-106)
			\Line[arrow,arrowpos=0.5,arrowlength=15,arrowwidth=10,arrowinset=1](320,-106)(480,-106)
			\Photon(360,-37)(360,-106){5}{5.5}
			\Photon(440,-37)(440,-106){5}{5.5}
			\end{picture}
			&
			\begin{picture}(226,182) (287,-197)
			\SetWidth{1.0}
			\SetColor{Black}
			\Arc[arrow,arrowpos=0.5,arrowlength=15,arrowwidth=10,arrowinset=1,clock](400,-106)(80,-180,-360)
			\Arc[arrow,arrowpos=0.5,arrowlength=15,arrowwidth=10,arrowinset=1](400,-106)(80,-180,0)
			\Line(288,-106)(320,-106)
			\Line(480,-106)(512,-106)
			\Line[arrow,arrowpos=0.5,arrowlength=15,arrowwidth=10,arrowinset=1](320,-106)(480,-106)
			\Photon(360,-175)(360,-106){5}{5.5}
			\Photon(440,-37)(440,-106){5}{5.5}
			\end{picture}
			&
			\begin{picture}(226,182) (287,-197)
			\SetWidth{1.0}
			\SetColor{Black}
			\Arc[arrow,arrowpos=0.5,arrowlength=15,arrowwidth=10,arrowinset=1,clock](400,-106)(80,-180,-360)
			\Arc[arrow,arrowpos=0.5,arrowlength=15,arrowwidth=10,arrowinset=1](400,-106)(80,-180,0)
			\Line(288,-106)(320,-106)
			\Line(480,-106)(512,-106)
			\Line[arrow,arrowpos=0.25,arrowlength=15,arrowwidth=10,arrowinset=1](320,-106)(480,-106)
			\PhotonArc(400.5,5.8)(68.802,-134.823,-45.177){5}{6.5}
			\Photon(400,-68)(400,-106){5}{2.5}
			\end{picture}
			&
			\begin{picture}(226,172) (287,-207)
			\SetWidth{1.0}
			\SetColor{Black}
			\Arc[arrow,arrowpos=0.2,arrowlength=15,arrowwidth=10,arrowinset=1,clock](400,-116)(80,-180,-360)
			\Arc[arrow,arrowpos=0.5,arrowlength=15,arrowwidth=10,arrowinset=1](400,-116)(80,-180,0)
			\Line(288,-116)(320,-116)
			\Line(480,-116)(512,-116)
			\Line[arrow,arrowpos=0.5,arrowlength=15,arrowwidth=10,arrowinset=1](320,-116)(480,-116)
			\PhotonArc(400.5,-28.8)(54.202,-153.482,-26.518){5}{7.5}
			\Photon(400,-77)(400,-36){5}{3.5}
			\end{picture}
			\\
			{\Huge $\left( s \right)$}
			&
			{\Huge $\left( t \right)$}
			&
			{\Huge $\left( u \right)$}
			&
			{\Huge $\left( v \right)$}
		\end{tabular*}
	}
	\caption{Four-loop diagrams contributing to the Yukawa-dependent part of the anomalous dimension}
	\label{diagfoura}
\end{table}
\clearpage

The simple pole contributions in Table~\ref{restaba} may now be summed using Eq.~\eqref{poles} and we obtain
the Yukawa-dependent contribution to the four-loop anomalous dimension using Eq.~\eqref{gamdef}.  
 Our final result is
 \begin{align}
 (8\pi)^4\gamma_{\Phi}^{(4)}=&\tfrac23\gamma^{(4)}_1+\tfrac{\pi^2}{4}\gamma^{(4)}_2
 +\tfrac43\left(1-\tfrac13\pi^2\right)\gamma^{(4)}_3-\left(4-\tfrac23\pi^2\right)\gamma^{(4)}_4\nn
 &-\tfrac13\pi^2(\gamma^{(4)}_7+\gamma^{(4)}_8)\nn
 &+\left[2\left(1-\tfrac18\pi^2\right) \Ttil
 +\tfrac18\pi^2C_G\right]\left(\tfrac13\gamma^{(4)}_9-\gamma^{(4)}_{10}\right)
-\tfrac43\gamma^{(4)}_{11}+\ldots,
 \label{finalres}
 \end{align}
where as before the ellipsis indicates Yukawa-independent terms. We may now read off the coefficients $g_i^{(4)}$ as defined in Eq.\eqref{gdef}, and in particular we see that $g_5^{(4)}=g_6^{(4)}=0$,
$g_7^{(4)}=g_8^{(4)}=-\tfrac13\pi^2$, in accord with Eq.~\eqref{fourcon}.

\begin{table}[t]
	\setlength{\extrarowheight}{1cm}
	\setlength{\tabcolsep}{24pt}
	\hspace*{-10.5cm}
	\centering
	\resizebox{6.7cm}{!}{
		\begin{tabular*}{20cm}{ccccc}
			\begin{picture}(244,162) (246,-207)
			\SetWidth{1.0}
			\SetColor{Black}
			\Arc(368,-126)(80,127,487)
			\Arc(245.902,-50.996)(122.2,-65.282,2.343)
			\Arc(490.098,-50.996)(122.2,177.657,245.282)
			\Arc(503.236,-208.502)(211.413,129.768,167.293)
			\Line(247,-162)(297,-162)
			\Line(439,-162)(489,-162)
			\end{picture}\hspace*{1cm}
			&
			\begin{picture}(162,122) (287,-287)
			\SetWidth{1.0}
			\SetColor{Black}
			\Arc(328,-206)(39.962,122,482)
			\Arc(408,-206)(39.962,122,482)
			\Line(248,-206)(488,-206)
			\end{picture}\hspace*{3cm}
			&
			\begin{picture}(162,162) (287,-207)
			\SetWidth{1.0}
			\SetColor{Black}
			\Arc(245.902,-50.996)(122.2,-65.282,2.343)
			\Arc(368,-126)(80,127,487)
			\Arc(490.098,-50.996)(122.2,177.657,245.282)
			\Arc[clock](368,-311.531)(165.531,115.399,64.601)
			\Line(247,-162)(297,-162)
			\Line(439,-162)(489,-162)
			\end{picture}\hspace*{2cm}
			&
			\rotatebox{90}{\begin{picture}(172,222) (277,-177)
				\SetWidth{1.0}
				\SetColor{Black}
				\Arc(368,-66)(80,127,487)
				\Line(368,44)(368,-176)
				\Arc(428,-66)(100,126.87,233.13)
				\Arc[clock](308,-66)(100,53.13,-53.13)
				\SetWidth{0.0}
				\Vertex(288,-66){9.899}
				\end{picture}}\hspace*{1cm}
			&
			\rotatebox{90}{\begin{picture}(182,222) (277,-177)
				\SetWidth{1.0}
				\SetColor{Black}
				\Line(305,-114)(357,-74)
				\Line(430,-114)(379,-74)
				\Line(305,-18)(357,-60)
				\Line(430,-16)(379,-60)
				\Line(368,44)(368,-176)
				\Arc[arrow,arrowpos=0.5,arrowlength=15,arrowwidth=10,arrowinset=1,clock](366.942,-61.932)(75.939,144.654,89.201)
				\Arc[arrow,arrowpos=0.5,arrowlength=15,arrowwidth=10,arrowinset=1,clock](368.648,-63.727)(77.73,90.478,37.88)
				\Arc[arrow,arrowpos=0.5,arrowlength=15,arrowwidth=10,arrowinset=1,clock](368.853,-66.268)(79.157,39.423,-37.085)
				\Arc[arrow,arrowpos=0.5,arrowlength=15,arrowwidth=10,arrowinset=1,clock](370,-70)(76.026,-35.362,-91.507)
				\Arc[arrow,arrowpos=0.5,arrowlength=15,arrowwidth=10,arrowinset=1,clock](369.75,-62.5)(83.518,-91.201,-141.929)
				\Arc[arrow,arrowpos=0.5,arrowlength=15,arrowwidth=10,arrowinset=1,clock](368,-66)(80,-143.13,-216.87)
				\end{picture}}
			\\
			\hspace*{-1cm}{\Huge $I_4$}
			&
			\hspace*{-2.5cm}{\Huge $I_{22}$}
			&
			\hspace*{-2cm}{\Huge $I_{4bbb}$}
			&
			\hspace*{-1cm}{\Huge $\tilde Y$}
			&
			\hspace*{0cm}{\Huge $X$}
		\end{tabular*}
	}
	\caption{The basis of massless two-point momentum integrals. For definitions and results, see Appendix B as well as the main text. The dotted propagator is squared, and the arrows on the last integral denote numerator factors of $\kslash$.}
	\label{diagfourc}
\end{table}
\begin{table}[h]
	\begin{center}
		\begin{tabular}{|c| c c c c c| c |}\hline
			&symm&$I_4$&$I_{22}$&$I_{4bbb}$&$\tilde Y$&overall group factor\\ [0.5ex]\hline
			(a)&$-\tfrac{1}{12}$&1&0&0&0&$\gamma^{(4)}_1$\\ [0.5ex]
			(b)&$-\tfrac{1}{8}$&0&0&1&0&$\gamma^{(4)}_2$\\ [0.5ex]
			(c)&$-\tfrac14$&$-2$&0&0&0&$\gamma^{(4)}_4-\tfrac{1}{4}C_G\gamma^{(4)}_{10}$\\ [0.5ex]
			(d)&$-\tfrac12$&$0$&0&$1$&0&$\gamma^{(4)}_4-\tfrac{1}{2}C_G\gamma^{(4)}_{10}$\\ [0.5ex]
			(e)&$-1$&$0$&$0$&$\tfrac23$&0&$-\tfrac12\gamma^{(4)}_4-\tfrac12\gamma^{(4)}_{12}-\tfrac{1}{12}C_G\gamma^{(4)}_9+\tfrac{1}{4}C_G\gamma^{(4)}_{10}
			$\\ [0.5ex]
			(f)&
			$1$&$0$&$0$&$-\tfrac23$&0&
			$\gamma^{(4)}_8+\tfrac{1}{12}C_G\gamma^{(4)}_9-\tfrac14C_G\gamma^{(4)}_{10}$\\ [0.5ex]
			(g)&
			$-\tfrac14$&0&0&$-2$&0&$\gamma^{(4)}_8+\tfrac{1}{12}C_G\gamma^{(4)}_9-\tfrac14C_G\gamma^{(4)}_{10}$\\ [0.5ex]
			(i)&$\tfrac12$&$0$&$0$&$-\tfrac23$&0&
			$\tfrac12\gamma^{(4)}_4+\tfrac12\gamma^{(4)}_7-\gamma^{(4)}_8+\gamma^{(4)}_{12}$\\ [0.5ex]
			(j)&$1$&$-2$&0&1&0&$\gamma^{(4)}_7-\tfrac{1}{12}C_G\gamma^{(4)}_9$\\ [0.5ex]
			(k)&$-\tfrac12$&$-2$&0&0&0&$\gamma^{(4)}_7-\tfrac{1}{12}C_G\gamma^{(4)}_9$\\ [0.5ex]
			(l)&$1$&$-2$&$0$&$\tfrac43$&0&$\tfrac16\gamma^{(4)}_3-\tfrac12\gamma^{(4)}_7$\\ [0.5ex]
			(m)&$-\tfrac{1}{12}$&$-2$&0&0&0&$\gamma^{(4)}_3-\tfrac{1}{4}C_G\gamma^{(4)}_9$\\ [0.5ex]
			(n)&
			$-\tfrac{1}{2}$&$1$&0&$-\tfrac12$&0&$\left(\tilde T+\tfrac12C_G\right)\left(\tfrac16\gamma^{(4)}_9-\tfrac12\gamma^{(4)}_{10}\right)$
			\\ [0.5ex]
			(o)&$-\tfrac{1}{2}$&$0$&$\tfrac12$&0&0&$\left(\tilde T+\tfrac12C_G\right)\gamma^{(4)}_{10}$\\ [0.5ex]
			(p)&$\tfrac{1}{2}$&$0$&$\tfrac12$&0&0&$\left(\tilde T+\tfrac12C_G\right)\gamma^{(4)}_{10}$\\ [0.5ex]
			(q)&$-\tfrac{1}{12}$&$0$&$1$&0&$-2$&$\gamma^{(4)}_{11}$\\ [0.5ex]
			(r)&$\tfrac{1}{6}$&$1$&$0$&0&$-1$&$\gamma^{(4)}_{11}$\\ \hline
		\end{tabular}
		\caption{Results for diagrams listed in Table~\ref{diagfoura} in terms of master integrals (see 
Table~\ref{diagfourc}) and invariants involving Yukawa couplings of Eq.~\eqref{Ystruct}}.
		\label{restaba}
	\end{center}
\end{table}

\bigskip

\section{Conclusions}

We have demonstrated the existence of an $a$-function having the gradient flow properties of Eq.~\eqref{grad} at next-to-leading order, for a general three-dimensional $\Ncal=2$ supersymmetric gauge theory in this paper; and for a completely general (non-supersymmetric) ungauged three-dimensional theory in Ref.~\cite{JPa}. It is worth emphasising that in our $a$-function construction we have had to compute a new class of Feynman diagram at four loops to ensure full consistency. For instance, we found by computation that graph $X$ of Table 6 was finite.
Alternatively we could have used properties of our $a$-function construction to have predicted this a priori. That the two approaches tally is indicative that our $a$-function and the field theories we have examined are fully informed of each other.
 It seems highly likely that the gradient flow property will extend to a completely general three-dimensional Chern-Simons theory coupled to scalars and fermions at next-to-leading order and probably beyond; and again, based on this one might obtain predictions for further Feynman diagrams which would otherwise require advanced techniques to evaluate.

It would be very desirable to find a general all-orders proof, or to make contact with the $F$-function described in 
Refs.~\cite{jaff,klebb,kleba} which has been argued to have similar properties at least at leading order. In this connection it might be interesting to compute the $a$-function as in Eqs.~\eqref{adef}, \eqref{aeqs}, \eqref{acoeffs} for the particular theories considered in Refs.~\cite{jaff, klebb} in order to make a direct comparison. It would be all the more desirable to relate our $a$-function with the $F$-function since as we mentioned earlier, it has been shown that the latter increases as expected between IR and UV fixed points; whereas although we have demonstrated monotonic behaviour of our $a$-function perturbatively, i.e. for weak couplings (since our metric is the unit matrix at leading order), we currently have no way to prove this in general.  

\subsection{Acknowledgments}
JAG and IJ were supported by the STFC under contract ST/G00062X/1, CP by an STFC studentship, 
and YS by FONDECYT project 1151281 and UBB project GI-152609/VC.

\begin{appendix}
\section{Conventions}

In this appendix we list our superspace and supersymmetry conventions.
We use a metric signature $(+--)$ so that a possible choice of $\gamma$ matrices
is $\gamma^0=\sigma_2$, $\gamma^1=i\sigma_3$, $\gamma^2=i\sigma_1$
with
\be
(\gamma^{0})_{\alpha}{}^{\beta}=(\sigma_2)_{\alpha}{}^{\beta},
\ee
etc. We then have
\be
\gamma^{\mu}\gamma^{\nu}=\eta^{\mu\nu}-i\epsilon^{\mu\nu\rho}\gamma_{\rho}.
\ee
We have\cite{penatib}
 a complex two-spinor $\theta^{\alpha}$ (with conjugate denoted $\thetabar^{\alpha}$)
with indices raised and lowered according to
\be
\theta^{\alpha}=C^{\alpha\beta}\theta_{\beta},\quad
\theta_{\alpha}=\theta^{\beta}C_{\beta\alpha},
\ee
with $C^{12}=-C_{12}=i$. We then have
\be
\theta_{\alpha}\theta_{\beta}=C_{\beta\alpha}\theta^2,\quad
\theta^{\alpha}\theta^{\beta}=C^{\beta\alpha}\theta^2,
\ee
where
\be
\theta^2=\tfrac12\theta^{\alpha}\theta_{\alpha}.
\ee
The supercovariant derivatives are defined by
\begin{align}
D_{\alpha}=&\pa_{\alpha}+\tfrac{i}{2}\thetabar^{\beta}\pa_{\alpha\beta},\\
\Dbar_{\alpha}=&\pabar_{\alpha}+\tfrac{i}{2}\theta^{\beta}\pa_{\alpha\beta},
\end{align}
where
\be
\pa_{\alpha\beta}=\pa_{\mu}(\gamma^{\mu})_{\alpha\beta},
\label{pdef}
\ee
satisfying
\be
\{D_{\alpha},\Dbar_{\beta}\}=i\pa_{\alpha\beta}.
\ee
We also define
\be
d^2\theta=\tfrac12d\theta^{\alpha}d\theta_{\alpha}
\quad d^2\thetabar=\tfrac12d\thetabar^{\alpha}d\thetabar_{\alpha},\quad 
d^4\theta=d^2\theta d^2\thetabar,
\ee
so that
\be
\int d^2\theta\theta^2=\int d^2\thetabar\thetabar^2=-1.
\ee
The vector superfield $V(x,\theta,\thetabar)$ is expanded in Wess-Zumino gauge
as
\be
V=i\theta^{\alpha}\thetabar_{\alpha}\sigma+\theta^{\alpha}\thetabar^{\beta}
A_{\alpha\beta}-\theta^2\thetabar^{\alpha}\lambdabar_{\alpha}
-\thetabar^{2}\theta^{\alpha}\lambda_{\alpha}+\theta^2\thetabar^{2}D,
\ee
and the chiral field is expanded as
\be
\Phi=\phi(y)+\theta^{\alpha}\psi_{\alpha}(y)-\theta^2F(y),
\ee
where
\be
y^{\mu}=x^{\mu}+i\theta\gamma^{\mu}\thetabar.
\ee

\section{Integrals}

Here, we list the UV divergences of our basis of momentum integrals.
As in Refs.~\cite{MSS,MSSa}, these are subdivergence-subtracted massless two-point functions, depicted schematically in Table 6.
The basic massless 1-loop integral (defined to be dimensionless here) is given by\footnote{The $G$-function notation was introduced in Ref.~\cite{kat}.}
\begin{eqnarray}
G(a,b) &=& \int\!\!\frac{d^dk}{(2\pi)^d}\,\frac{p^{2(a+b-d/2)}}{k^{2a}\,(k-p)^{2b}}
\;=\;\frac{\Gamma\left(a+b-\tfrac{d}2\right)\,\Gamma\left(\tfrac{d}2-a\right)\,\Gamma\left(\tfrac{d}2-b\right)}
{(4\pi)^{d/2}\Gamma(d-a-b)\,\Gamma(a)\,\Gamma(b)}\,.
\end{eqnarray}
A standard method is then to iteratively integrate out massless sub-graphs in higher-loop integrals in terms of the function $G$, and this is indeed sufficient to evaluate the first four of the integrals of Table 6.
Due to dimensional reasons, in odd dimensions the first logarithmic UV divergence can only occur at even loop orders, which here is parametrized by the 2-loop massless sunset-type integral \begin{eqnarray}
I_2 &=& \int\!\!\frac{d^dk}{(2\pi)^d} \int\!\!\frac{d^dq}{(2\pi)^d}\, \frac{p^{2(3-d)}}{k^2\,q^2\,(k+q-p)^2}
\;=\; G(1,1)\,G\left(2-\tfrac{d}{2},1\right)\;.
\end{eqnarray}
\newcommand{\pole}[1]{\hat{K}\left[#1\right]}
We are now ready to define our basis of momentum integrals, using an operator $\pole{f(\epsilon)}$ that extracts the pole parts of the function $f(\epsilon)$. 
Recalling that we work in $d=3-\epsilon$ dimensions, we obtain \begin{eqnarray}
I_4 &=& \pole{I_2\left(G(1,1)\,G\left(5-\tfrac{3d}{2},1\right)-\pole{I_2}\right)}
= \frac1{(8\pi)^4}\left(-\frac2{\epsilon^2}+\frac4{\epsilon}\right)\,,\nn
I_{22} &=& \pole{I_2\left(I_2-2\pole{I_2}\right)}
= \frac1{(8\pi)^4}\left(-\frac4{\epsilon^2}+\frac0{\epsilon}\right)\,,\nn
I_{4bbb} &=& \pole{G^3(1,1)\,G\left(4-d,2-\tfrac{d}{2}\right)} = \frac1{(8\pi)^4}\left(\frac{\pi^2}{\epsilon}\right)\,,\nn
\tilde{Y} &=& \pole{G^2(1,1)\,G\left(2,2-\tfrac{d}{2}\right)\,G\left(2-\tfrac{d}{2},4-d\right)}
= \frac1{(8\pi)^4}\left(-\frac2{\epsilon}\right)\,,
\label{poles}
\end{eqnarray}
the first three of which agree with Refs.~\cite{MSS,MSSa} after adjusting for our definition of $\epsilon$ which differs by a factor of $2$. 
Furthermore, the integrals satisfy the consistency condition $4\tilde{Y}=I_{22}-2I_4$ given in Eq.~(4.1) of Ref.~\cite{JPa}. We note that in fact the result for $\tilde Y$ is not required since it cancels between rows (q) and (r) of Table~\ref{restaba}.

\end{appendix}

\end{document}